\documentclass[11pt]{article}
\usepackage[margin=1in]{geometry}
\usepackage[utf8]{inputenc}
\usepackage{amsmath}
\usepackage{amssymb}
\usepackage{amsthm}
\usepackage{float}
\usepackage{algorithm}
\usepackage{algpseudocode}
\usepackage{graphicx}
\usepackage{hyperref}
\usepackage{url}
\usepackage{cite}
\usepackage{tabularx}
\usepackage{booktabs}
\usepackage{xfrac}
\usepackage{nicefrac}
\usepackage{tikz}
\usetikzlibrary{
    arrows.meta,
    shapes.geometric,
    positioning
}
\newcommand{\lastaccessed}{\newline Last accessed: March \number\year}
\theoremstyle{definition}
\newtheorem{definition}{Definition}[section]
\newtheorem{guarantee}{Guarantee}[section]
\newtheorem{property}{Property}[section]
\newtheorem{example}{Example}[section]

\title{Bitcoin Smart Accounts: Trust-Minimized\\Native Bitcoin DeFi Infrastructure}
\author{Cian Lalor \\Lombard Finance \and Matthew Marshall\\Lombard Finance \and Antonio Russo\\Lombard Finance}
\date{\today}

\begin{document}

\maketitle

\begin{abstract}
Bitcoin's limited programmability and transaction throughput have historically prevented native Bitcoin from participating in decentralized finance (DeFi) applications. Existing solutions depend on honest-majority thresholds, or centralized custodial entities that introduce significant trust requirements. This paper introduces Bitcoin Smart Accounts (BSA), a novel protocol that enables native Bitcoin to access DeFi through trust-minimized infrastructure while maintaining self-custody of funds.

BSA achieves this through a combination of emulated Bitcoin covenants using Partially Signed Bitcoin Transactions (PSBTs) and Taproot scripts, a Trusted Execution Environment (TEE)-based arbitration system, and destination chain smart contracts that enable DeFi platforms to accept self-custodial Bitcoin as collateral without necessitating protocol-level modifications. The setup leverages liquidity secured by the Lombard Security Consortium which provides a twofold advantage: for a DeFi protocol, liquidators rely on fungible assets with deep liquidity to quickly exit positions, while for a depositor, the general trust assumptions of honest majority ($m$-of-$n$) are reduced to existential honesty ($1$-of-$k$).

We present the complete protocol design, including the Bitcoin architecture, the TEE-based arbitration mechanism, and the Smart Account Registry for protocol management. We provide a security analysis that demonstrates the correctness, safety, and availability properties under our trust model. Our design enables native Bitcoin to serve as collateral in lending markets and other DeFi protocols without requiring users to relinquish custody of funds.

\end{abstract}

\section{Introduction}

Bitcoin~\cite{bitcoin} has emerged as the dominant cryptocurrency by market capitalization, with more than \$1 trillion in value locked in the network. However, Bitcoin's conservative design philosophy, which prioritizes security and decentralization, has resulted in limited transaction throughput and programmability. These constraints have prevented Bitcoin from participating in the rapidly growing decentralized finance (DeFi) ecosystem, where programmable smart contracts enable lending, borrowing, trading, and other financial primitives.

\subsection{The Problem}

To access DeFi functionality, Bitcoin holders have historically faced a fundamental trade-off: either trust a custodial service to hold their Bitcoin to issue fungible tokens, or use bridges that require trusting an honest majority of operators ($m$-of-$n$ threshold schemes). Both approaches introduce significant risk that undermines Bitcoin's core value proposition of self-custody and trust minimization.

Observed limitations of existing bridges and Bitcoin Layer 2 solutions include:
\begin{itemize}
    \item \textbf{Custodial Risk:} Most wrapped Bitcoin solutions require users to deposit funds into centralized custody, creating single points of trust and failure, introducing excessive counterparty risk.
    \item \textbf{Trust Assumptions:} Threshold signature schemes require trusting that a majority of signers remain honest, which highlights an attack vector for adversarial actors. Current implementations regularly rely on complex key management (adding or removing signers, key rotation).
    \item \textbf{Novel Cryptography:} Recent advances in Bitcoin bridging solutions are based on succinct knowledge proofs implemented over garbled circuits~\cite{babylon-vaults}. These cryptographic primitives are novel, and complex to audit and guarantee correctness.
    \item \textbf{Limited Programmability:} Bitcoin's intentionally limited scripting language, and present lack of native covenant support, prevent the enforcement of complex spending conditions directly on the Bitcoin network.
    \item \textbf{Integration Challenges:} Existing solutions often create friction for DeFi protocols, involving liquidity delays and bespoke infrastructure for the Bitcoin network, particularly around liquidation mechanisms that are critical for lending markets, e.g., Babylon Bitcoin Vaults~\cite{babylon-vaults}.
\end{itemize}

\subsection{Our Contribution}

This paper introduces \textbf{Bitcoin Smart Accounts (BSA)}, a protocol built using existing and well-established primitives that enables native Bitcoin to access DeFi applications while maintaining self-custody and minimizing trust assumptions. Our contributions include:

\begin{enumerate}
    \item \textbf{Trust-Minimized Security Model:} We reduce the depositor's (the user bridging Bitcoin) trust assumptions from an honest majority ($m$-of-$n$) to existential honesty ($1$-of-$k$), where a depositor requires a minimum of only one other honest party in the protocol to be safe.
    
    \item \textbf{TEE-Based Arbitration Oracle:} We introduce a Trusted Execution Environment (TEE)-based arbitration system (Sec.~\ref{sec:arbitration}) that verifies cross-chain events and resolves disputes. The TEE ensures code integrity and provides cryptographic attestations of correct execution, enabling the arbitration logic to be executed by any third party.

    \item \textbf{Emulated Covenant Architecture:} We design a Bitcoin script architecture using Taproot addresses~\cite{taproot} and Partially Signed Bitcoin Transactions (PSBTs) to emulate covenant-like functionality, i.e., transaction constraints enforced with a committee, without requiring Bitcoin consensus changes—an approach also used in Lightning Network~\cite{lightning-network} and BitVM~\cite{bitvm2-bridge}. This enables complex, multi-party spending conditions to be enforced without relying on a trusted third party.
    
    \item \textbf{DeFi-Native Integration:} We design the Smart Account Registry, a smart contract that enables seamless integration with existing DeFi protocols, supporting standard liquidation mechanisms without creating impediments for liquidators or lending protocols by leveraging existing on-chain liquidity through the use of Lombard's existing Security Consortium (a collection of validators requiring $\nicefrac{2}{3}$ consensus—see footnote marker \textit{*} on Table.~\ref{tab:comparison} for further detail). 
    
    \item \textbf{Security Analysis:} We provide security evaluations that demonstrate the safety, liveness, and integrity properties under our trust model (Sec. \ref{sec:security}).
\end{enumerate}

\subsection{BSA Comparison to Existing Bridging Solutions}

Trustless approaches such as BitVM~\cite{bitvm2-bridge} achieve the strongest theoretical security guarantees by verifying arbitrary computations on Bitcoin through fraud proofs. However, they introduce significant operational complexity, including multi-round interactive protocols, large on-chain footprints, and requirements for DeFi applications to integrate directly with the Bitcoin network, making them impractical for general-purpose DeFi use cases today. BSA takes a pragmatic approach: by combining emulated covenants with TEE-based arbitration, it achieves comparable security to fully trustless vaults, specifically $1$-of-$k$ existential honesty and self-custody, while being deployable today with significantly lower operational complexity and native compatibility with existing DeFi infrastructure.

Below, we provide a table~\ref{tab:comparison} to compare the key properties of our contribution with related works.

\begin{table}[h]
\centering
\footnotesize
\setlength{\tabcolsep}{4pt}
\newcolumntype{L}[1]{>{\raggedright\arraybackslash}p{#1}}
\newcolumntype{C}[1]{>{\centering\arraybackslash}p{#1}}
\begin{tabular}{|L{2.8cm}|L{2.5cm}|L{2.5cm}|L{2.5cm}|L{2.8cm}|}
\hline
\textbf{Property} & \textbf{Bitcoin Smart Accounts} & \textbf{Multisig Bridge} & \textbf{BitVM Bridge} & \textbf{Centralized Issuer (wrapped BTC)} \\
\hline
Trust Model & $1$-of-$k$ & $m$-of-$n$* & $1$-of-$k$ & $1$-of-$1$ \\
\hline
Self-Custody & Yes & No & Yes & No \\
\hline
Native DeFi support (e.g., liquidations) & Yes, with programmable perimeter† & Yes & No, requires DeFi applications to integrate with Bitcoin network & Yes \\
\hline
Decentralized withdrawal of native BTC by depositor & Yes, assumes $1$-of-$(k+1)$‡ honest. & Yes, assumes $m$-of-$n$ honest. & Yes, assumes $1$-of-$k$ honest. & No. Requires permissions from centralized authority. \\
\hline
Required Operators to steal BTC from depositor & $k$-of-$k$ offline AOs and dishonest TO‡ & $m$-of-$n$  operators & $k$-of-$k$ operators & 1 single operator \\
\hline
Withdrawal of native BTC by non-depositor & Indirect via TO rebalance & Yes & Indirect via Operator & Yes \\
\hline
\end{tabular}
\caption{Comparison of Bitcoin Smart Accounts with related work.}
\label{tab:comparison}
\vspace{-0.2cm}
\begin{flushleft}
\textit{* Lombard's Security Consortium (the TO), a multisig bridge, includes additional safeguards to reduce the trust model further, such as the integration of Cubist's Bascule}~\cite{bascule-drawbridge} \textit{and Chainlink oracles}~\cite{chainlink-attestation} \textit{to provide a secondary validation for all fund movements during normal operation.}\\
\textit{†The BSA enables seamless liquidations without liquidator- or DeFi- customizations. For more information, see section~\ref{sec:liquidation-process}.}\\
\textit{‡An AO cannot be dishonest, only offline. When considering attack scenarios, we view the protocol to have $1$-of-$(k+1)$ security, given that the protocol consists of $k$ AOs and a single TO. This is explained in more detail in section~\ref{subsec:trust-model}.}
\end{flushleft}
\end{table}

\subsection{Paper Organization}
The remainder of this paper is organized as follows. Section~\ref{sec:model} presents our system model, assumptions, and security goals. Section~\ref{sec:bitcoin-arch} details the complete protocol design, including the Bitcoin script architecture, PSBT mechanisms, and state transitions. Section~\ref{sec:onchain} covers the Smart Account Registry and the components on the destination chain. Section~\ref{sec:arbitration} describes the Arbitration Oracle TEE implementation. Section~\ref{sec:security} provides security analysis. Section~\ref{sec:limitations} discusses limitations and future work. Section~\ref{sec:conclusion} concludes. The appendices outline additional key implementation details.

\section{System Model and Security Goals}
\label{sec:model}

\subsection{Network Model}

We assume that Bitcoin and the destination chain (e.g., Ethereum) operate as robust public ledgers with persistence and high-availability properties. We make standard cryptographic assumptions: participants are computationally bounded, and standardized hash functions, signatures, and encryption schemes are utilized. 

\subsection{Participants}

The BSA protocol involves the following participants:

\begin{itemize}
    \item \textbf{Depositor ($D$):} A user who wants to utilize their native Bitcoin within DeFi applications. The depositor maintains control of their Bitcoin and destination chain private keys and participates in the Protocol User Setup Ceremony.
    
    \item \textbf{Token Operator ($TO$):} A single acting party responsible for minting BTC.b\footnote{BTC.b is a Bitcoin derivative minted by the Lombard Security Consortium, which is backed one-to-one by native Bitcoin.} and executing Bitcoin rebalances e.g., for DeFi liquidations.  In practice, this is Lombard's existing Security Consortium, consisting of the existing set of validators that secures Lombard through $2/3$ majority consensus. This actor can be viewed as the \emph{canonical operator} for protocol instantiation: a requirement for an instance of the BSA to exist.
    
    \item \textbf{Arbitration Oracles ($AO$):} A software module that verifies protocol states and resolves disputes. A set of $k$ arbitration oracles may exist, and the depositor requires at least one honest arbitration oracle to avoid trusting only the TO. The $k$ AOs are interchangeable for a given protocol instance: any correct AO (see below Def.~\ref{def:ao-correctness}) included in the depositor's BSA instance setup can resolve its disputes.
\end{itemize}

\subsection{Failure Model}
\label{sec:failure-model}
The depositor and the Token Operator are Byzantine parties assumed to be selfish and rational, i.e. they will do anything which allows them to gain an economic advantage, and they do not harm themselves.

The Arbitration Oracle may fail due to lack of availability, i.e. it may not fulfill time-sensitive duties. However, it cannot act outside the protocol's design constraints or produce inaccurate results. Multiple independent entities can play the Arbitration Oracle role in the same protocol instance in such a way that the Arbitration Oracle duties are fulfilled as long as at least one performs within the protocol time constraints.

\begin{definition}[AO Correctness]
\label{def:ao-correctness}
An Arbitration Oracle is \textit{correct} if it is online and operational, i.e., it is available to verify protocol states and sign transactions resolving disputes in favor of the depositor, according to the deterministic dispute resolution logic as outlined in Sec.~\ref{sec:tee}.
\end{definition}

An AO operator (entity running the AO software) may be malicious, but due to the TEE-enforced code integrity (Sec.~\ref{gar:tee-code-integ}) and emulated covenant constraints (Sec.~\ref{sec:psbt-convenant-emulation}), the most severe adversarial action available to a malicious AO operator is to take the AO offline. A malicious AO operator cannot sign incorrect transactions or redirect funds to unauthorized addresses, as all spending paths are constrained by pre-signed PSBTs to commit to specific output addresses.

Whenever a party behaves according to the protocol rules, it is said to be \emph{correct}. For the AO, correctness is equivalent to availability, as a malicious AO operator cannot violate protocol rules due to TEE and covenant constraints.

\subsection{Security Goals}
\label{sec:security-goals}
We define the security properties of the protocol according to the perspective of each participating party.

\begin{definition}[Protocol Liveness]
    \label{def:protocol-liveness}
    A protocol is said to have \emph{liveness} if funds cannot be prevented from exiting the protocol after they are transferred to the address VA, as defined in Sec. ~\ref{sec:script-addresses}.
\end{definition}

\begin{definition}[Depositor Safety]
    \label{def:depositor-safety}
    A protocol instance is \emph{Depositor Safe} if there exists a finite and time bounded list of operations that allows a correct depositor to regain full and unilateral control of its funds on the Bitcoin blockchain.
\end{definition}

\begin{definition}[Token Operator Safety]
    \label{def:token-operator-safety} 
    A protocol instance is \emph{Token Operator Safe} if in any protocol state where the corresponding token collateralized by Bitcoin deposits is minted on the destination chain, there exists a finite and time-bounded list of operations that makes the Token Operator gain full and unilateral control of funds on the Bitcoin blockchain when collateral tokens exit the protocol perimeter as defined in Sec.~\ref{sec:perimeter}.
\end{definition}

\begin{definition}[Protocol Safety]
    \label{def:protocol-safety}
    A protocol instance is \emph{Protocol Safe} if and only if both Depositor Safety \ref{def:depositor-safety} and Token Operator Safety \ref{def:token-operator-safety} hold.
\end{definition}

Given the Failure Model described in section \ref{sec:failure-model}, the general security goal of this work is to guarantee Protocol Safety~\ref{def:protocol-safety} under the assumption that at least $1$-of-$k$ entities playing the Arbitration Oracle role are correct.

As a consequence, if the TO is malicious, the depositor must trust that at least $1$-of-$k$ AOs remain honest. This is a significant improvement over traditional $m$-of-$n$ threshold schemes.

The protocol description following in the next section will make apparent that from the depositor's point of view, for $k$ AOs, the $1$-of-$k$ requirement reduces to $1$-of-$(k+1)$ since the honest behavior of the TO does not require any intervention of AO.

\section{Bitcoin Architecture}
\label{sec:bitcoin-arch}

\subsection{Overview}

The BSA protocol enables a depositor to lock native Bitcoin in a self-custodial vault while receiving BTC.b receipt tokens on a destination chain for use in DeFi applications. The protocol uses four Taproot script addresses with specific spending conditions, enforced through pre-signed PSBTs exchanged during the \emph{Protocol User Setup Ceremony} (Sec. \ref{subsubsec:setup-cermony}).

\subsection{Protocol Parameters}
\label{sec:protocol-parameters}

The BSA protocol uses the following time lock parameters:

\begin{itemize}
    \item \textbf{$T_1$}: The relative time lock period during which the TO can challenge a withdrawal, or unbonding, request. After $T_1$ expires without challenge by the TO, the depositor can unilaterally claim funds.
    
    \item \textbf{$T_2$}: The relative time lock period for any AO to resolve a state in favor of the depositor. After expiration, the funds can be pulled by the TO.

    \item \textbf{$T_3$}: The relative time lock period for any core protocol modifications to take effect (e.g., upgrading the destination chain smart contracts), which may impact the security of the protocol. Hence, all security guarantees (see Sec.~\ref{sec:sec-guarantees}) of the protocol are upheld until $T_3$.
\end{itemize}

where the time lock relation:
\begin{equation}
\label{eq:timelocks-relation}
    T_3 > (T_1 + T_2)
\end{equation}
 ensures correctness. The depositor and an AO can verify this relation before depositing Bitcoin.

\subsection{Bitcoin Script Architecture}
We design four Taproot script addresses~\cite{taproot} that form the core architecture on the Bitcoin blockchain. Taproot scripts were chosen for their reduced transaction size and fees, and ability to enable smart contract-like spending logic.

Every user has their own suite of unique script addresses, directly tied to both their destination chain address and Bitcoin network public key. The usage of PSBTs exchanged by the parties allows funds to be directed between the addresses of the protocol in a verifiable manner. Verifiability is achieved through a public code repository which outlines formulae for address derivation. Appendix~\ref{app:address-tweaking} provides more details on the topic.

\subsubsection{Script Addresses}
\label{sec:script-addresses}
The four Taproot script addresses represent states within the protocol. Funds move through these addresses according to the protocol rules. A user's deposit may contain multiple UTXOs, which can be held at any of a user's set of unique addresses while the protocol is live. An instance of the protocol is considered to be live if these UTXOs are locked to any one of the four addresses:

\begin{enumerate}
    \item \textbf{Vault Address (VA):} a $2$-of-$2$ multisig between a single depositor and the TO. This is where the depositor initially locks their Bitcoin.
    \item \textbf{Unbond Timelock Address (UTA):} a $2$-of-$2$ multisig between a single depositor and TO, with an additional $1$-of-$1$ spending path for the depositor after the timelock $T_1$ expires.
    \item \textbf{Unbond Challenge Address (UCA):} a $2$-of-$2$ multisig between a single depositor and a set of AOs, with an additional $1$-of-$1$ spending path for the TO after timelock $T_2$ expires.
    \item \textbf{Rebalance Challenge Address (RCA):} a $2$-of-$2$ multisig between a single depositor and a set of AOs, with an additional $1$-of-$1$ spending path for the TO after timelock $T_2$ expires.
\end{enumerate}

In order to map transactions to corresponding state transitions, each taproot address has its internal key tweaked by an address-specific label. In this way, each of the listed Bitcoin addresses is unique with respect to an instance of the protocol.

The internal key of all Taproot addresses is a Nothing Up My Sleeves (NUMS) key, according to BIP-0341~\cite{taproot}. See Appendix~\ref{app:address-tweaking} for more information.

\begin{figure}
    \centering
    \begin{tikzpicture}[
        font=\small,
        box/.style={
            draw,
            rounded corners,
            minimum width=3.2cm,
            minimum height=1.4cm,
            align=center
        },
        actor/.style={
            draw,
            minimum width=3.2cm,
            minimum height=1.4cm,
            align=center
        },
        arrow/.style={->, thick},
        dashedarrow/.style={->, thick, dashed}
    ]
    
    
    \coordinate (top)        at (0, 6.5);
    \coordinate (vaultc)     at (0, 4.8);
    
    \coordinate (leftc)      at (-5, 2.5);
    \coordinate (rightc)     at ( 5, 2.5);
    
    \coordinate (midc)       at (0, 0.0);
    
    \coordinate (bottoml)    at (-5, -2.5);
    \coordinate (bottomr)    at ( 5, -2.5);

    
    \node (utxo) at (top) {1. user deposited UTXOs};
    
    \node[box] (vault) at (vaultc) {Vault Address};
    
    \node[box] (rebalance) at (leftc) {Rebalance\\Challenge\\Address};
    
    \node[box] (timelock) at (rightc) {Unbond\\Timelock\\Address};
    
    \node[draw, circle, minimum size=2cm, align=center] (depositor) at (midc) {Depositor};
    
    \node[draw, circle, minimum size=2cm, align=center] (operator) at (bottoml) {Token\\Operator};
    
    \node[box] (unbond) at (bottomr) {Unbond\\Challenge\\Address};
    
    
    \draw[arrow] (utxo) -- (vault);
    
    
    \draw[dashedarrow]
        (vault.south) -- node[right=4pt, align=center] {7. Cooperative\\Unbond} (depositor.north);
    
    
    \draw[dashedarrow]
        (vault.west) -| node[midway, above] {5. Rebalance Request} (rebalance.north);
    
    \draw[dashedarrow]
        (vault.east) -| node[midway, above] {2a. Unbond} (timelock.north);
    
    
    \draw[dashedarrow]
        (rebalance.south east) -- node[above left, xshift=50pt, yshift=10pt] {6a. AO resolve} (depositor.150);
    
    \draw[arrow]
        (rebalance.south) -- node[left, align=center] {6b.\\$T_2$ expiry} (operator.north);
    
    
    \draw[arrow]
        (timelock.south west) -- node[above right, xshift=-10pt, yshift=-22pt] {2b. $T_1$ expiry} (depositor.30);
    
    \draw[dashedarrow]
        (timelock.south) -- node[right, align=center] {3. Unbond\\Challenge} (unbond.north);
    
    
    \draw[dashedarrow]
        (unbond.north west) -- node[above left, xshift=65pt] {4a. AO resolve} (depositor.330);
    
    \draw[arrow]
        (unbond.west) -- ++(-5,0)
        node[pos=0.66, below] {4b. $T_2$ expiry}
        |- (operator.east);
    
    \end{tikzpicture}
    \caption{State transition diagram showing the vault unbond paths. We use boxes to represent addresses with spending paths by two or more members, and circles to represent addresses controlled by an individual actor. Terms in diagram reference state transitions as enumerated in section ~\ref{subsec:state-transitions}. Executors of state transitions can be seen in section ~\ref{subsubsec:setup-cermony}. Dotted lines reference 2-of-2 spending paths, solid lines reference 1-of-1 spending paths. }
    \label{fig:state-transitions}
\end{figure}
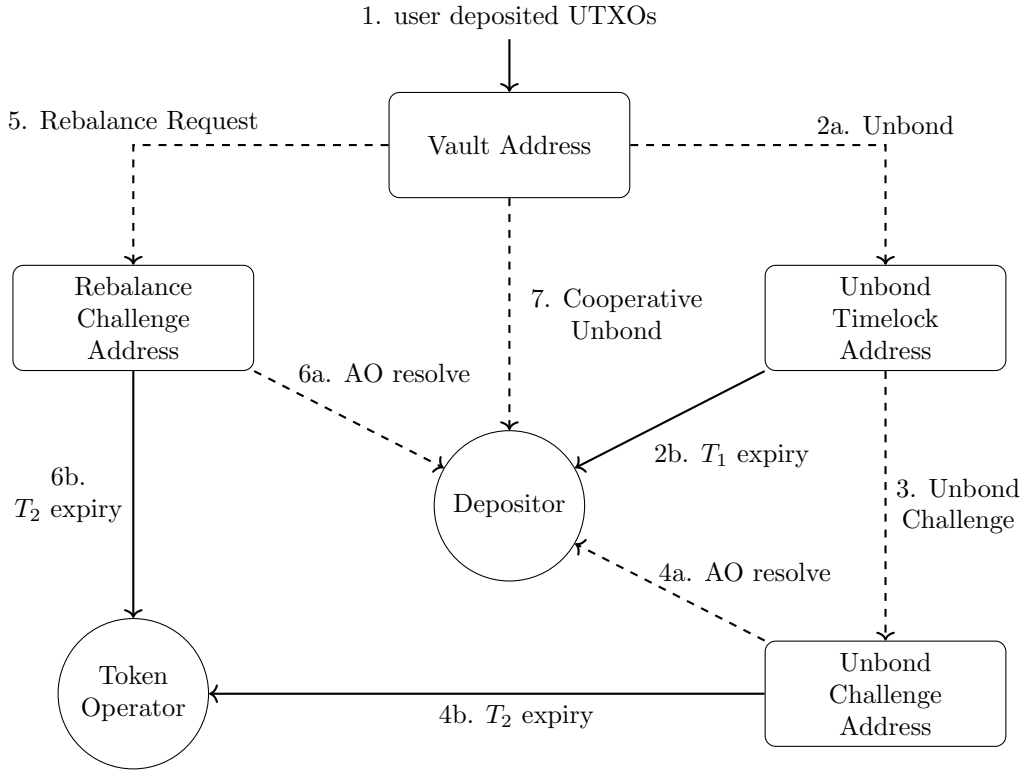

\subsection{PSBT-Based Covenant Emulation}
\label{sec:psbt-convenant-emulation}
Bitcoin Script does not natively support covenants (constraints on transaction outputs). Native covenant support would require opcodes such as \texttt{OP\_CHECKTEMPLATEVERIFY}~\cite{bip119} that enable scripts to inspect and constrain transaction outputs. We emulate covenant functionality by combining multi-signature spending constraints with pre-signed transactions that commit to specific outputs. During Protocol User Setup Ceremony (Sec.~\ref{subsubsec:setup-cermony}), the TO, AO, and depositor exchange a set of PSBTs. These PSBTs commit to specific spending paths and output addresses, effectively constraining future transaction outputs. The key insight is that by pre-signing transactions with fixed outputs, we can enforce spending conditions that would otherwise require native covenant support. This approach builds upon the techniques used by other protocols, e.g., the Lightning Network~\cite{lightning-network}. 

A covenant emulation is established between two parties: the party who creates and provides a PSBT (the \emph{creator}), which pre-commits transaction outputs to specific destination addresses, and the party who receives the PSBT (the \emph{executor}), who can sign and execute the transaction using the inputs and outputs specified in the PSBT received by the creator.

In a valid protocol instance, all parties have signed and provided the expected PSBTs. Addresses in PSBT outputs commit to the expected spending scripts.

\subsubsection{Protocol User Setup Ceremony}
\label{subsubsec:setup-cermony}
Bitcoin's UTXO model enables precise control over which specific UTXOs are spent. For each UTXO deposited into the VA, the following set of PSBTs are created and exchanged for the \emph{Protocol User Setup Ceremony} to start a valid protocol instance:

\begin{enumerate}
    \item The depositor receives a PSBT from the TO enabling them to initiate an unbonding request (moving funds from VA to UTA) unilaterally.
    \item The TO receives PSBTs from the depositor, enabling them to unilaterally:
    \begin{enumerate}
        \item Challenge depositor unbonding requests (moving funds from UTA to UCA).
        \item Initiate rebalancing requests (moving funds from VA to RCA).
    \end{enumerate}
    \item The AO receives PSBTs from the depositor enabling them to unilaterally resolve:
    \begin{enumerate}
        \item Challenged unbonding requests in favor of the depositor (moving funds from UCA to Dep).
        \item Rebalance requests in favor of the depositor (moving funds from RCA to Dep).
    \end{enumerate}
\end{enumerate}

The number of PSBTs required from the depositor can be reduced through optimization techniques, such as having the TO provide PSBTs for certain spending paths, though this extends trust assumptions. For a complete overview of transactions, please refer to the table in Appendix~\ref{app:tx-reference}.

\begin{figure}[]
    \centering
    \begin{tikzpicture}[
        font=\normalsize,
        node distance=.7cm,
        sbox/.style={draw, rounded corners, minimum height=0.6cm, align=left, fill=white, text width=11.5cm, inner sep=5pt},
        arr/.style={-{Stealth[length=3mm, width=2mm]}, thick},
        phaselabel/.style={font=\small\bfseries, gray!80}
    ]

    \node[sbox] (s0) {\textbf{0.} Depositor shares intended UTXOs to fund VA with TO};
    \node[sbox, below=of s0] (s1) {\textbf{1a.} TO signs: VA\textrightarrow{}UTA \\ \textbf{1b.} Dep signs: VA\textrightarrow{}RCA, UTA\textrightarrow{}UCA, RCA\textrightarrow{}Dep, UCA\textrightarrow{}Dep};
    \node[sbox, below=of s1] (s2) {\textbf{2a.} TO verifies: VA\textrightarrow{}RCA, UTA\textrightarrow{}UCA \\\textbf{2b.} TO posts on SAR: VA\textrightarrow{}UTA, UCA\textrightarrow{}Dep, RCA\textrightarrow{}Dep};
    \node[sbox, below=of s2] (s3) {\textbf{3.} Dep verifies on SAR: VA\textrightarrow{}UTA, UCA\textrightarrow{}Dep, RCA\textrightarrow{}Dep};
    \node[sbox, below=of s3] (s4) {\textbf{4.} Depositor transfers BTC to VA};
    \node[sbox, below=of s4] (s5) {\textbf{5.} TO mints BTC.b after 6 confirmations on Bitcoin network};

    \draw[arr] (s0) -- (s1);
    \draw[arr] (s1) -- (s2);
    \draw[arr] (s2) -- (s3);
    \draw[arr] (s3) -- (s4);
    \draw[arr] (s4) -- (s5);

    \node[phaselabel, anchor=west] at ([xshift=0.3cm]s0.east) {Identify};
    \node[phaselabel, anchor=west] at ([xshift=0.3cm]s1.east) {Sign};
    \node[phaselabel, anchor=west] at ([xshift=0.3cm]s2.east) {\shortstack [l] {TO Verify \\ \& Store}};
    \node[phaselabel, anchor=west] at ([xshift=0.3cm]s3.east) {Dep. Verify};
    \node[phaselabel, anchor=west] at ([xshift=0.3cm]s4.east) {Deposit};
    \node[phaselabel, anchor=west] at ([xshift=0.3cm]s5.east) {Mint};

    \end{tikzpicture}
    \caption{Protocol User Setup Ceremony. PSBTs are exchanged and signed, posted to the SAR (Sec.~\ref{sec:onchain}), and independently verified before the depositor funds the Vault Address and the TO mints BTC.b at the corresponding destination address.}
    \label{fig:setup-ceremony}
\end{figure}
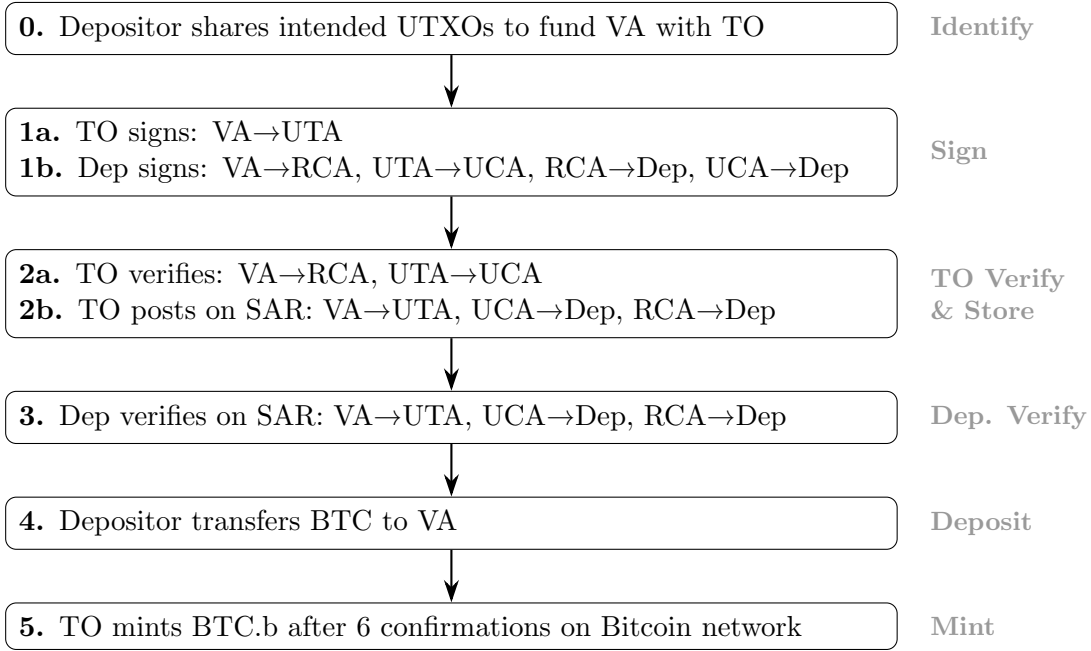

\subsubsection{Dynamic PSBT Updates}

While the initial PSBT exchange occurs during protocol setup, additional PSBTs can be added to the protocol while it is live to optimize workflow and enable new functionalities, such as a fast unbond from the Vault in case the TO cooperates and the depositor behaves correctly (see \textbf{Cooperative Unbond} below).

\subsection{State Transitions}
\label{subsec:state-transitions}
The protocol supports the following state transitions:

\begin{enumerate}
    \item \textbf{Deposit:} Depositor sends Bitcoin to VA. TO observes the deposit, and if protocol is considered valid according to section \ref{sec:psbt-convenant-emulation}, mints the corresponding BTC.b on the destination chain.
    
    \item \textbf{Unbond Request:} The depositor burns BTC.b on the destination chain and then (a) initiates an unbonding transaction from VA to UTA, and if the TO does not challenge, then (b) after time lock $T_1$ expires, the depositor pulls the funds from UTA.
    
    \item \textbf{Unbond Challenge:} If TO observes an unbonding request but no corresponding burn event, TO challenges by moving funds from UTA to UCA.
    
    \item \textbf{Unbond Challenge Resolution:} AO verifies the destination chain state and either: (a) resolves the dispute in favor of the depositor, moving funds from UCA to the depositor's address, or (b) chooses inaction, effectively resolving the dispute in favor of the TO, enabling the TO to pull funds from UCA after $T_2$.
    
    \item \textbf{Rebalance Request:} TO observes an under-collateralized position and triggers a rebalance event on the destination chain, moving funds from VA to RCA.
    
    \item \textbf{Rebalance Finalization:} AO verifies the destination chain state and either: (a) resolves the rebalance request in favor of the depositor, moving funds from RCA to the depositor's address, or (b) chooses inaction, effectively finalizing the rebalance in favor of the TO, enabling the TO to pull funds from RCA after $T_2$.

    \item \textbf{Cooperative Unbond:} \textit{this state transition is not strictly necessary for protocol functionality, and is introduced only to improve user experience.} If a user requests the TO to verify that they may legitimately unbond (i.e., have already burnt BTC.b), the TO will sign an additional PSBT enabling a spending path from the vault directly to the depositor's return address, circumventing the required time lock for manual unbonds. 
    
    \textit{Note: There is no further reference to Cooperative Unbond in this paper as it does not impact participants' trust or protocol security.}
\end{enumerate}

As outlined in section ~\ref{sec:script-addresses}, a depositor's instance of the protocol contains four unique Bitcoin addresses. A depositor can deposit multiple UTXOs to an instance of the BSA, effectively splitting funds across UTXOs which can move independently across protocol states. Splitting funds is beneficial for rebalance and unbond mechanisms, e.g., enabling partial liquidations (see Sec.~\ref{sec:utxo-rebalancing}).

\begin{figure}[H]
    \centering
    \makebox[\textwidth][c]{%
    \begin{tikzpicture}[
        font=\footnotesize,
        xscale=1.0,
        event/.style={draw, rounded corners, fill=white, minimum width=1.4cm, minimum height=0.55cm, align=center, inner sep=2pt, font=\footnotesize},
        arr/.style={-{Stealth[length=2mm, width=1.5mm]}, semithick},
        chainlbl/.style={font=\footnotesize\bfseries}
    ]

    \def\xSetup{0}
    \def\xDeposit{2}
    \def\xMint{3.2}
    \def\xDeFi{5.6}
    \def\xBurn{7.8}
    \def\xUnbond{9}
    \def\xChallenge{12}
    \def\xResolve{14.7}

    \def\yBtc{1.0}
    \def\yDest{-1.7}
    \def\yActorBtc{0.4}
    \def\yActorDest{-1.15}

    \node[chainlbl] at (-1.5, 0.5) {Bitcoin};
    \node[chainlbl] at (-1.5, -1.3) {Dest.};

    \draw[gray!40, semithick] (-0.3, -0.4) -- (15.5, -0.4);

    \foreach \x/\t in {\xSetup/$t_0$, \xDeposit/$t_1$, \xMint/$t_2$, \xDeFi/$t_3$, \xBurn/$t_4$, \xUnbond/$t_5$, \xChallenge/$t_6$, \xResolve/$t_7$} {
        \draw[gray!40] (\x, -0.2) -- (\x, -0.6);
        \node[below, gray!60, font=\tiny] at (\x, -0.6) {\t};
    }

    \node[event] (setup) at (\xSetup, \yBtc) {Setup};
    \node[event] (btcVA) at (\xDeposit, \yBtc) {BTC in VA};
    \node[event] (vaUta) at (\xUnbond, \yBtc) {VA$\to$UTA};
    \node[event] (utaUca) at (\xChallenge, \yBtc) {UTA$\to$UCA};
    \node[event] (resolved) at (\xResolve, \yBtc) {UCA$\to$Dep};

    \node[event] (mint) at (\xMint, \yDest) {BTC.b minted};
    \node[event] (defi) at (\xDeFi, \yDest) {DeFi};
    \node[event] (burn) at (\xBurn, \yDest) {Burn BTC.b};

    \draw[arr] (setup) -- (btcVA);
    \draw[arr] (btcVA) -- node[right, font=\tiny, align=center, pos=0.4] {6 confs} (mint);
    \draw[arr] (mint) -- (defi);
    \draw[arr] (defi) -- (burn);
    \draw[arr] (burn) -- (vaUta);
    \draw[arr] (vaUta) -- node[above, font=\tiny, align=center] {unfair\\challenge} (utaUca);
    \draw[arr] (utaUca) -- node[above, font=\tiny, align=center] {AO\\resolve} (resolved);

    \node[font=\tiny, blue!70] at (\xSetup, \yActorBtc) {Dep};
    \node[font=\tiny, blue!70] at (\xDeposit, \yActorBtc) {Dep};
    \node[font=\tiny, red!70] at (\xMint, \yActorDest) {TO};
    \node[font=\tiny, blue!70] at (\xDeFi, \yActorDest) {Dep};
    \node[font=\tiny, blue!70] at (\xBurn, \yActorDest) {Dep};
    \node[font=\tiny, blue!70] at (\xUnbond, \yActorBtc) {Dep};
    \node[font=\tiny, red!70] at (\xChallenge, \yActorBtc) {TO};
    \node[font=\tiny, green!50!black] at (\xResolve, \yActorBtc) {AO};

    \end{tikzpicture}%
    }
    \caption{Protocol lifecycle timeline: onboarding through to offboarding with an illegitimate challenge.}
    \label{fig:lifecycle}
\end{figure}
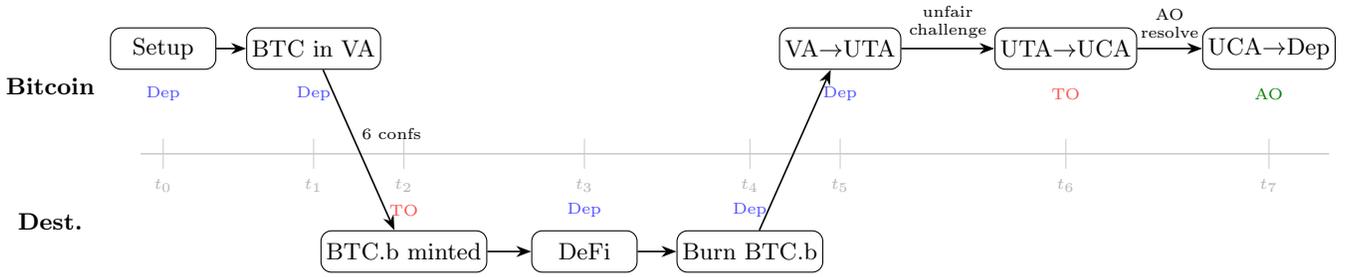

\subsection{Rebalance Models}

When the TO is honest, the \textit{Rebalance Request} is a reactionary state transition triggered by the TO whenever an arbitrary amount of BTC.b tokens exits the protocol perimeter as defined in Sec.~\ref{sec:perimeter}. Given this amount may be arbitrary, it will likely not correspond to the specific UTXO amounts locked in the VA for a given user's instance of the BSA.

To overcome this misalignment, we propose two rebalancing models to balance user experience and protocol flexibility. 

\textit{Note: When the TO is dishonest, this state transition will be corrected by an AO (see Sec.~\ref{gar:depositor-safety}).}

\subsubsection{Model 1: Collaborative Rebalancing}
\label{sec:coll-rebalancing}

A user deposits a single UTXO in the VA. When a rebalance is required, the TO requests a new set of PSBTs from the depositor within a time window, enabling new protocol states that reflect the amount to be rebalanced from the original deposit and the remaining amount. If the depositor fails to sign within the agreed-upon time window, the TO rebalances the full amount of the deposited UTXO to the RCA. 

\subsubsection{Model 2: UTXO-Based Rebalancing}
\label{sec:utxo-rebalancing}

A user transfers multiple UTXOs to the same VA, and a corresponding amount of tokens are minted to the destination address. The number of UTXOs determines the granularity of rebalancing. When a rebalance occurs, the TO moves sufficient UTXOs from VA to RCA to cover the imbalance. As PSBTs, signed during the Protocol User Setup Ceremony, commit to amounts in deposited UTXOs, it is not feasible to partially liquidate any whole UTXOs. \\

Importantly, in both models, for any over-seized amount, the depositors' DeFi position(s) remains active. The only effective change is that part of the underlying collateral backing said position is transferred from the depositor's self-custodial BSA vault to the LSC's general reserves: any amount of over-seized collateral does not impact the fungibility of the corresponding on-chain BTC.b. Any amount remaining in the VA is unchanged, and still fully collateralizes a DeFi position as before a rebalance request. Consequently, when the depositor wishes to unbond they need to claim any over-seized amount from the TO in the event the required rebalance does not match the user's exact deposited amounts.

\subsection{Fee Management}

In the BSA protocol, without proper fee management, UTXOs moving through state transitions would "consume themselves" to pay for transaction costs. However, as PSBTs are pre-signed during the Protocol User Setup Ceremony, and since these signatures commit to both inputs and outputs, fees cannot be updated if network conditions change and the preset fees become insufficient. This introduces a problem: if a pre-signed transaction's fee becomes too low relative to current network conditions, the UTXO could become stuck and the protocol would not progress.

To prevent the protocol from being unable to execute a state transition, we use a combination of \textit{Anchor Outputs}~\cite{anchor-outputs} and \textit{Child-Pays-For-Parent (CPFP)} transactions~\cite{cpfp}, along with \textit{SigHash} flags ~\cite{taproot}. 

Both the mechanisms outlined below in the remainder of this section allow the protocol to avoid committing to unnecessarily high fees during the Protocol User Setup Ceremony, while ensuring transactions can always be executed and confirmed, independent of network conditions.

\subsubsection{Anchor and CPFP Transactions}
For all state transitions involving a covenant emulation between the TO and the depositor, the \emph{creator} includes an additional anchor output which the \emph{executor} can spend unilaterally.

If a pre-signed transaction $tx_0$ is broadcast but not confirmed due to insufficient fees ($f_0 < f_{\text{required}}$), the executor can construct a CPFP transaction $tx_{\text{anchor}}$ that spends both the anchor output from the unconfirmed $tx_0$ and an additional UTXO controlled by the executor. By setting the fee of $tx_{\text{anchor}}$ such that $f_{\text{anchor}} \geq f_{\text{required}} + \tilde{f}_{\text{required}} - f_0$, where $\tilde{f}_{\text{required}}$ is the additional required fee for $tx_\text{anchor}$,  the total effective fee for $tx_0$ becomes,
\begin{equation}    
f_{\text{total}} = f_0 + f_{\text{anchor}} \geq f_{\text{required}} + \tilde{f}_{\text{required}},
\end{equation}
ensuring that $tx_0$ is confirmed.

\subsubsection{Sighash Additional UTXO}
For AO-invoked transitions where the AO does not commit to future protocol state, we adopt transaction signatures that specify the flag \texttt{SIGHASH\_ANYONECANPAY | SIGHASH\_ALL} instead of adding Anchor outputs to reduce protocol costs. These flags are set during the Protocol User Setup Ceremony when the depositor signs the PSBT for the AO's resolutions. This flag enables any entity, e.g. the AO, to add an additional UTXO as an input, which can be used to pay higher fees during network congestion.

If a pre-signed transaction $tx_0$ is broadcast but not confirmed due to insufficient fees ($f_0 < f_{\text{required}}$), the executor can add an additional input as a fee $f_1$, to $tx_0$ ensuring that $f_{\text{total}} = f_0 + f_{1} \geq f_{\text{required}}$.\\

\section{Smart Account Registry}
\label{sec:onchain}

\subsection{Overview}

The Smart Account Registry (SAR) is a smart contract deployed on the destination chain (e.g. Ethereum) that serves as the central coordination point for the protocol. The SAR is responsible for:
\begin{itemize}
    \item Maintaining a public registry of deposit UTXOs and protocol parameters. This includes a status for every deposited UTXO, used by the AO to determine the correct resolution of any dispute (Sec.~\ref{sec:tee}). 
    \item Defining the \textit{perimeter} (Sec. \ref{sec:perimeter}) and triggering events when depositor imbalances are detected, legitimizing Bitcoin rebalancing (Sec.\ref{sec:rebalance-mechanism}).
    \item Defining the rebalancing order of UTXOs, set by the depositor, described in section \ref{sec:rebalance-order}.
    \item Storing of pre-signed PSBTs for data availability, ensuring they cannot be corrupted or modified after protocol setup
    \item Storing authorized AO software versions with corresponding expiry timestamps, signed by the TO, (Sec.\ref{subsec:TEE-upgrades}).
\end{itemize}

\subsection{Rebalance}

A rebalance is the process when tokens minted on the destination chain against BTC held in an instance of the BSA exit the \textit{perimeter} (as defined in Sec.~\ref{sec:perimeter}) requiring the native Bitcoin acting as collateral to be transferred to the Lombard Security Consortium's general reserves. The token minted on the destination chain can be burnt and redeemed for native BTC from these general reserves after exiting the perimeter, while funds held in the BSA can only be transferred to predefined states as outlined in the Protocol User Setup Ceremony (Sec.~\ref{subsubsec:setup-cermony}). As redemptions are resolved using BTC from the general reserve, the rebalance mechanism is used to pull Bitcoin from BSA instances into the general reserves in the event said instance of the BSA is undercollateralized, rectifying any depositor imbalance. 

\subsubsection{Rebalance Mechanism}
\label{sec:rebalance-mechanism}
 For the state transition \emph{Rebalance Request} on the Bitcoin network to be legitimate, it must be preceded by a rebalance event on the destination chain, which is a result of a user \emph{imbalance}.  An imbalance occurs when the sum of initially deposited UTXOs $D$ is greater than the sum of the user's balances across all DeFi platforms and their personal address.

Formally, let $B_{\text{defi}} = \sum_{i=1}^{k} b_i$ denote the aggregate balance across $k$ DeFi platforms, and $B_{\text{personal}}$ denote the personal address balance, we let the total,

\begin{equation}
B_{\text{total}} = B_{\text{defi}} + B_{\text{personal}}
\end{equation}

and an imbalance is detected when,

\begin{equation}
B_{\text{total}} < D
\end{equation}

When an imbalance is detected, the TO can then submit a transaction to the Smart Account Registry to mark the depositor as requiring rebalancing, emitting an event from the smart contract. This event allows the TO to initiate a legitimate rebalance request for the depositor's underlying Bitcoin without the risk of AO intervention. \\

\subsubsection{Perimeter Definition}
\label{sec:perimeter}
The \textit{perimeter} is defined as an additional set of smart contracts registered on the SAR, called \textit{Adapters}, which collectively track the value of $B_\text{total}$. These adapters enable the efficient querying of $B_\text{total}$ for each user. This design choice was motivated to help both facilitate integration with DeFi protocols and allow BTC.b to be simultaneously backed by Bitcoin in BSA protocol and the TO's general reserves, without redemption risk to any BTC.b holder. 

Adapter management follows a conservative security model: adding new adapters (integrating a new DeFi protocol) is effective immediately, as adapters only return positive or zero balances, posing no risk to depositors. However, removing or replacing an adapter requires adherence to time lock $T_3$, allowing depositors sufficient time to manage their DeFi positions in the event they are affected.

\subsubsection{UTXO Rebalance Ordering}
\label{sec:rebalance-order}
Every user address has an ordered set of UTXO outputs,  $O = [o_1, \ldots, o_n]$, where each $o_i \in O$ is an active deposit on the SAR for any $i$. Users can customize the rebalance ordering by defining a permutation function $\pi: [n] \to [n]$, where $[n] = (1, \dots, n)$, mapping initial output indices to the desired rebalance order, returning a sequence $L = [o_{\pi(1)}, o_{\pi(2)}, \ldots, o_{\pi(n)}]$ where $L[1]$ is rebalanced first.

\subsection{Rebalancing Process}
\label{sec:liquidation-process}
When a rebalance is detected, as defined in section ~\ref{sec:rebalance-mechanism}, the TO initiates the process of rebalancing, sending from either $\text{VA} \rightarrow \text{RCA}$, or $\text{UTA} \rightarrow \text{UCA}$ if a depositor's illegitimate unbond request was confirmed on the Bitcoin network. The rebalancing takes at most $T_2$ to complete, after which the TO can pull the funds into Lombard's general reserves.

For the sake of clarity, we stress that the derivative token BTC.b, issued by the TO (Security Consortium)\footnote{Lombard's Security Consortium's general reserves operate as a permissionless bridge: depositors generate a unique deterministic Bitcoin deposit address using a set of parameters including destination chain and address, and BTC.b is minted on the corresponding chain and address.}, is backed by both BTC held in Lombard's general reserves and BSA instances.

\begin{example}[Liquidation on a Lending Protocol]
Suppose a liquidation occurs on a lending protocol from an unhealthy position that uses BTC held in a BSA instance as collateral. A liquidator repays the user's bad debt (borrowed asset) to the lending platform, and as a result, acquires the collateral for this position, BTC.b, backed by native BTC from a user's  BSA instance\footnote{In practice, a subset of DeFi platforms will rely on third-party oracles to track the reserves of BTC.b, ensuring that collateral is tracked regardless of whether funds are held in individual BSA instances or in Lombard's general reserve. In this scenario, the reserve oracles track and attest that addresses for every instance of the BSA are included in this list, which a depositor can verify before depositing BTC.}. This causes the user's BSA to become imbalanced (see Sec. ~\ref{sec:rebalance-mechanism}) as their BTC.b has exited the perimeter. 

After receiving BTC.b from a liquidation, the liquidator has two options: (1) swap or sell the BTC.b on a decentralized exchange (DEX), or (2) redeem the BTC.b one-to-one for native Bitcoin from Lombard's general reserve.

In the second case, the redemption time will vary based on the proportion of liquidity immediately available in Lombard's general reserves. The longest time this can take is when Lombard's general reserve contains no Bitcoin. The redemption will then be completed within at most $T_2+\omega$, where $\omega$ is the (nominal) time required for the TO to execute the transaction to pull the funds into its general reserves and transfer them to the liquidator's specified Bitcoin address. The address is provided as an input parameter by the liquidator\footnote{Incidentally, as the BSA protocol is agnostic to liquidations and liquidators, i.e., the liquidator is not required to update the SAR and mark a position as \texttt{Imbalanced} in order to retrieve the collateral, the user has a theoretical time window to restore their position by purchasing BTC.b on the open market before the imbalance is recorded by the TO. The TO, however, can check imbalances arbitrarily, so this may not be feasible in every scenario.} (required for a standard redemption) when calling the \texttt{redeem} function on the BTC.b contract.
\end{example}

UTXOs are rebalanced according to rebalancing order $L$ (as described in Sec.~\ref{sec:rebalance-order}) to cover the amount of imbalance  $\Delta = D - (B_{\text{defi}} + B_{\text{personal}})$. The minimal set of UTXOs from $L$ needed to cover $\Delta$ is selected for rebalance, with the remaining UTXOs unaffected. As mentioned in section ~\ref{sec:utxo-rebalancing}, as PSBTs commit to set amounts, in the event a user's deposited UTXO amounts cannot exactly cover the required rebalance, an over-seizure will occur where the TO will rebalance an amount sufficiently large to cover the rebalance. This over-seized amount will (1) still actively collateralize the user's DeFi position without interruption and (2) can be redeemed by the user during unbonding from the TO. A user can mitigate this concern by increasing the liquidation granularity by depositing more, and smaller, UTXOs.

\subsection{Upgrades}
\label{sec:sar-upgrades}
To maintain the ability to introduce new features and optimize protocol design in the future, the SAR implementation can be upgraded. The time lock for any proposed change is $T_3$, providing sufficient time for review by all participants in the protocol, and for participants to exit the protocol in the event they do not agree with scheduled changes.

\section{Arbitration Oracle Architecture}
\label{sec:arbitration}
\subsection{Overview}
The Arbitration Oracle (AO) plays a key role in ensuring that the Smart Account Registry (SAR) and the relevant protocol aspects for the Bitcoin network have consistent states. The AO is analogous to a neutral arbitrating party in a tri-party agreement. 

For an instance of the protocol to be live, the Enclave codebase, proposed by the TO, must be accepted by both the depositor and the AO operator. Firstly, the TO signs a specific source code and configuration for an AO version. Secondly, the AO reviews this version and can accept it by deploying it. Following the deployment of this AO version, a cryptographic attestation is made available by the AO to both the depositor and the TO, providing a guarantee of AO integrity (see below Guarantee~\ref{gar:tee-code-integ}). Lastly, based on this attestation, the TO accepts this specific AO deployment as a legitimate AO party (making the AO Bitcoin signing key eligible to partake in instances of BSA), and the depositor can review that the core logic in the enclave follows a set of codified arbitration rules to which they agree. 

\subsection{Trusted Execution Environment}
The main characteristic of the AO is that it can only fail due to lack of availability (offline), as described in the Failure Model (Sec. \ref{sec:failure-model}), i.e., the AO cannot sign maliciously outside of logic detailed in section \ref{sec:tee}. This guarantee is enforced by having all operations that validate protocol states and sign transactions deployed inside a hardware enforced trusted execution environment (TEE).

TEEs are technologies provided by major CPU vendors that enable the creation of execution environments with hardware-enforced isolation from other software running on the same platform. These environments can produce a cryptographic measurement of their initial code and configuration, and generate an attestation report authenticating this measurement. This procedure uses an attestation key rooted in hardware and certified by the vendor’s public key infrastructure.

By pairing such an attestation report with a key management service capable of verifying the report and conditioning key release on the measured code identity, an AO can be constrained to execute only the logic described in section \ref{sec:tee}, even if the operator hosting the AO behaves adversarially with respect to other protocol participants.

Our initial implementation is designed for the cloud provider Amazon Web Services (AWS). AWS provides a solution by integrating AWS Key Management Service (KMS)~\cite{aws-kms} with AWS Nitro Enclaves, a TEE solution based on the Firecracker~\cite{firecracker} virtualization technology. For an overview of how key management is implemented with AWS for an AO, refer to Appendix~\ref{app:arb-key}.

\subsection{Enclave Architecture}
\label{sec:enclave}
The enclave implements the core logic of the AO. This includes the list of logical operations that brings the AO to a protocol decision. This application is made up of two main components:
\begin{itemize}
    \item \textbf{Arbitration Logic:} Core implementation of the dispute resolution enforcing protocol rules 
    \item \textbf{Helios Light Client:} Implementation of the Altair Light Client protocol~\cite{altair} for SAR state verification.
\end{itemize}

The Enclave is packaged as an Enclave Image Format (EIF) file containing both the arbitration logic and the light client. The AO software package is built in a reproducible manner, so any resulting EIF always yields the same PCR0 measurement; participants can thus verify that the attested enclave matches the published source. Please see Appendix~\ref{app:repro-build} for an overview of implementation details.

The Enclave generates an AWS KMS key with policy restrictions that deny access outside the Enclave application, publishes the public key with AWS attestation including Platform Configuration Registries (PCRs) values that identify the exact codebase, and is limited to providing Bitcoin network signatures only under conditions enforced by the Enclave logic. Please see Appendix~\ref{app:kms-key-constraints} for an overview of implementation details.

\subsection{TEE Logic}
\label{sec:tee}
The aims of the TEE logic are to ensure the AO (i) processes only valid and relevant Bitcoin transactions, (ii) identifies both Bitcoin and SAR protocol states, and (iii) decides on a resolution based on the consistency or inconsistency of these states.

The AO operates in a fail-closed mode: it will only sign transactions (see figure ~\ref{fig:state-transitions}, labels 4a. \& 6a.) if all verification steps outlined below are completed successfully. If the AO cannot verify the SAR state (e.g., due to light client desynchronization, Sec.~\ref{subsec:light-client-sync}), it cannot sign. This design prioritizes safety over \emph{liveness} (see Def.~\ref{def:protocol-liveness}): no incorrect signatures can be produced under ambiguous conditions, thus, an AO that cannot verify is operationally equivalent to an offline AO. The depositor's liveness guarantee is preserved by the $1$-of-$k$ redundancy model. That is, a single correct AO with a synchronized view of the destination chain and legitimate inputs (Bitcoin transactions) is sufficient to resolve disputes within $T_2$.

\subsubsection{Transaction Verification}
\label{subsec:trans-verify}
If the enclave receives raw transaction data for $tx_{\text{VA}\rightarrow\text{RCA}}$; or $tx_{\text{VA}\rightarrow\text{UTA}}$ and $tx_{\text{UTA}\rightarrow\text{UCA}}$, the enclave: 

\begin{enumerate}
    \item verifies signatures match the public keys in the Bitcoin transaction's witness stack according to the UTXO value fetched from the SAR,
    \item verifies that one public key is the TO's public key $pk_{\text{TO}}$ (fetched from the SAR),
    \item verifies the address tweaking parameters (fetched from the SAR) and recognizes the state transition as described at the very end of section~\ref{sec:script-addresses}.
\end{enumerate}

For $tx_{\text{VA}\rightarrow\text{RCA}}$, the enclave:
\begin{enumerate}
    \setcounter{enumi}{3}
    \item identifies $\text{UTXO}_{\text{deposit}}$ from the input of $tx_{\text{VA}\rightarrow\text{RCA}}$.
\end{enumerate}

Or, for $tx_{\text{UTA}\rightarrow\text{UCA}}$, the enclave:
\begin{enumerate}
    \setcounter{enumi}{4}
    \item verifies the output of $tx_{\text{VA}\rightarrow\text{UTA}}$ is the input of $tx_{\text{UTA}\rightarrow\text{UCA}}$,
    \item identifies $\text{UTXO}_{\text{deposit}}$ as the input of $tx_{\text{VA}\rightarrow\text{UTA}}$.
\end{enumerate}

Lastly, the enclave queries the light client to:
\begin{enumerate}
    \setcounter{enumi}{6}
    \item identify the user account on the SAR which is indexed against $\text{UTXO}_{\text{deposit}}$,
    \item identify the SAR UTXO status,
    \item \label{step:9-trans-verify}verify the software the AO is running is up-to-date (see~\ref{sec:version-expiration} for details).
\end{enumerate}

The enclave then decides the appropriate action based on the logical rules outlined in sections ~\ref{RebalanceDisputeLogic} and~\ref{UnbondingDisputeResolutionLogic}, using the UTXO status to determine the correct resolution. 

\subsubsection{Rebalance Resolution}
\label{RebalanceDisputeLogic}
For rebalancing requests by the TO ($tx_{\text{VA}\rightarrow\text{RCA}}$), after completing the verification steps above, the Enclave checks the UTXO status on the SAR:
\begin{itemize}
    \item If the UTXO status is not \texttt{SpentOnRebalance} (indicating an invalid rebalance request by the TO), the enclave retrieves the PSBT from the SAR that enables the AO to resolve the dispute in favor of the depositor and signs it.
    \item Otherwise, the Enclave elects to perform no action, effectively resolving the dispute in favor of the TO, who can pull the funds after time lock $T_2$ expires.
\end{itemize}

\begin{algorithm}[H]
\caption{AO Enclave Logic for Rebalance Request Resolution.}
\label{alg:liquidation-rebalance}
\begin{algorithmic}[1]
\Require $tx_{\text{VA}\rightarrow\text{RCA}},  \text{sk}_{AO}$ (secret key from AO)
\State $\text{UTXO}_{deposit} \leftarrow tx_{\text{VA}\rightarrow\text{RCA}}.\text{input}$
\State $pk_{TO} \leftarrow \text{GetPKTOFromSAR}(\text{UTXO}_{deposit}$)
\State \textbf{if} $!\text{SigValid}(tx_{\text{VA}\rightarrow\text{RCA}}, pk_{TO})$ \textbf{then return} $\bot$
\State \textbf{if} $tx_{\text{VA}\rightarrow\text{RCA}}.output.address \neq \text{GetRCAFromSAR}(\text{UTXO}_{deposit})$ \textbf{then return} $\bot$
\State $utxoStatus \leftarrow \text{GetUTXOStatusFromSAR}(\text{UTXO}_{deposit})$
\State \textbf{if} $utxoStatus \neq SpentOnRebalance$ \textbf{then}
\State \quad $PSBT_{\text{depositor}} \leftarrow \text{LookupPSBTonSAR}(\text{UTXO}_{deposit})$ \Comment{PSBT to resolve in favour of depositor}
\State \quad $tx_{\text{resolved}} \leftarrow \text{Sign}(PSBT_{\text{depositor}}, \text{sk}_{\text{AO}})$
\State \quad \textbf{return} $tx_{\text{resolved}}$ 
\State \textbf{else}
    \State \quad \textbf{return} $\bot$ \Comment{Do nothing: rebalance authorized, TO can claim after $T_2$}
\end{algorithmic}
\end{algorithm}

\subsubsection{Unbonding Challenge Resolution}
\label{UnbondingDisputeResolutionLogic}
For challenged unbonding requests ($tx_{\text{UTA}\rightarrow\text{UCA}}$), after completing the verification of section~\ref{sec:tee}:
\begin{itemize}
    \item If the UTXO status is \texttt{Withdrawn} (indicating the depositor has burned BTC.b tokens) or \texttt{Rejected} (indicating the BTC.b was not yet minted and the depositor requested to exit the protocol), the Enclave retrieves the PSBT from the SAR that enables resolving the dispute in favor of the depositor and signs it.
    \item Otherwise, the Enclave elects to perform no action, effectively resolving the dispute in favor of the TO, who can pull the funds after time lock $T_2$ expires.
\end{itemize}

\begin{algorithm}[H]
\caption{AO Enclave Logic for Unbonding Challenge Resolution}
\label{alg:unbonding-dispute}
\begin{algorithmic}[1]
\Require $tx_{\text{UTA}\rightarrow\text{UCA}}$ and $tx_{\text{VA}\rightarrow\text{UTA}},  \text{sk}_{AO}$ (secret key from AO)
\State $\text{UTXO}_{deposit} \leftarrow tx_{\text{VA}\rightarrow\text{UTA}}.\text{input}$
\State $pk_{TO} \leftarrow \text{GetPKTOFromSAR}(\text{UTXO}_{deposit}$)
\State \textbf{if} $!\text{SigValid}(tx_{\text{VA}\rightarrow\text{UTA}}, pk_{TO})$ \textbf{then return} $\bot$
\State \textbf{if} $!\text{SigValid}(tx_{\text{UTA}\rightarrow\text{UCA}}, pk_{TO})$ \textbf{then return} $\bot$
\State \textbf{if} $tx_{\text{VA}\rightarrow\text{UTA}}.output \neq tx_{\text{UTA}\rightarrow\text{UCA}}.input$ \textbf{then return} $\bot$
\State \textbf{if} $tx_{\text{VA}\rightarrow\text{UTA}}.output.address \neq \text{GetUTAFromSAR}(\text{UTXO}_{deposit})$ \textbf{then return} $\bot$
\State \textbf{if} $tx_{\text{UTA}\rightarrow\text{UCA}}.output.address \neq \text{GetUCAFromSAR}(\text{UTXO}_{deposit})$ \textbf{then return} $\bot$
    \State $utxoStatus \leftarrow \text{GetUTXOStatusFromSAR}(\text{UTXO}_{deposit})$
    \State \textbf{if} $utxoStatus \in \{Withdrawn, Rejected\}$ \textbf{then}
    \State \quad $PSBT_{\text{depositor}} \leftarrow \text{LookupPSBTonSAR}(\text{UTXO}_{deposit})$ \Comment{PSBT to resolve in favour of depositor}
    \State \quad $tx_{\text{resolved}} \leftarrow \text{Sign}(PSBT_{\text{depositor}}, \text{sk}_{\text{AO}})$
    \State \quad \textbf{return} $tx_{\text{resolved}}$
    \State \textbf{else}
    \State \quad \textbf{return} $\bot$ \Comment{Do nothing: unbonding challenge authorized, TO can claim after $T_2$}
\end{algorithmic}
\end{algorithm}

\subsection{Light Client Synchronization}
\label{subsec:light-client-sync}
The TEE runs a light client of the destination chain in order to verify SAR events. As the first implementation of the protocol targets the Ethereum blockchain, the following discussion describes mechanics and constraints according to the Ethereum post-merge \cite{eth-merge} consensus protocol. Hence, depending on the destination chain, considerations for implementing BSA protocol roles will differ.

In Ethereum, the process of synchronizing with the network is subject to a weak subjectivity model~\cite{wsp}, where a Weak Subjectivity Period (WSP) $T_{sub}$ defines the maximum time a node can be offline while still resynchronizing autonomously. We stress the fact that an AO can resync autonomously to the network after both (1) a network interruption and (2) system downtime, both points detailed in Appendix~\ref{app:eth-consensus}. Period $T_{sub}$ is initialized with a conservative hardcoded value\footnote{A value is set to be shorter than any reasonable presumption for a WSP.} and, after the light client is synchronized, the TEE keeps updating the period length based on network conditions following the Ethereum specification.

When the TEE comes back online after downtime, it follows a resynchronization protocol: if downtime is less than $T_{sub}$, it resynchronizes from its last known block using a self-attested checkpoint \footnote{self-attested means that the TEE periodically, or upon request (e.g., in preparation for a scheduled reboot), creates an attestation as per section~\ref{sec:attestation} containing the currently computed WSP which is stored in a persistent storage by the entity running the enclave application}; if the downtime exceeds $T_{sub}$ or during initial boot, it uses a TO-signed checkpoint not older than the default hardcoded weak subjectivity period.

When an AO resynchronizes using a TO-provided checkpoint, they are responsible for verifying the validity of the checkpoint to ensure depositor safety. An AO can decline to resync if they disagree with the proposed checkpoint statement. An AO that declines to resync defaults to exiting the protocol until such a further time when it agrees with the TO's attested checkpoint. This is operationally equivalent to an offline AO. If any of a depositor's chosen AOs disagrees with a checkpoint, the depositor must be notified. The depositor can decide to maintain their BSA instance with their remaining AOs, or can choose to exit the protocol with sufficient time, as ensured by relation \ref{eq:timelocks-relation}. We explore how this impacts AO availability requirements in Appendix~\ref{app:avail-req}.

\subsection{Upgrades}
\label{subsec:TEE-upgrades}
The Enclave application can be upgraded to introduce new features, fix bugs, and for general code maintenance according to the following mechanics:
\begin{enumerate}
    \item Enclave Image Files (EIF) are signed by the TO.
    \item The AWS KMS key that decrypts the AO secret key is constrained via the PCR8 value, providing a guarantee from the Nitro Enclave system that only artifacts signed by the TO can operate it (i.e., the TO is the only entity able to release new software upgrades).
    \item AO providers are responsible for deploying and running such software in their cloud accounts; therefore, before updating the old image with the latest upgrade, they are responsible for verifying the contents of each upgrade.
\end{enumerate}
The above points ensure all protocol parties verify and approve the AO software while guaranteeing that only agreed software can use the AO KMS key. Depositors and the TO can request to verify the expected software is operated by the AO using the deterministic nature of Enclave Image Files builds and the PCR0 value reported in any attestation document. See Appendix~\ref{app:repro-build} for info on PCR values and Appendix~\ref{app:ver-lifecycle} for additional consideration on the AO version lifecycle.

\subsubsection{Version Expiration}
\label{sec:version-expiration}
Authorizing and verifying the new AO software does not provide additional security without the ability to deprecate previous versions. 

We achieve deprecation using a time-based expiration mechanism. The Smart Account Registry holds a mapping between versions and their respective expiration. The TO can set the expiration for a specific version by referring to its PCR0 value. The Enclave application, before performing any protocol action, verifies that its own version is not expired by retrieving the expiration value from the SAR. If the expiration date is past, the AO is not able to proceed with signing a transaction for dispute resolution (as shown in Sec. \ref{subsec:trans-verify} step \ref{step:9-trans-verify}.).

Due to the time-based expiration mechanism, AO operators have a time window within which they can update the Enclave without any downtime. The protocol parameter relation~\ref{eq:timelocks-relation} is enforced by the AO operator deploying the initial versions of the AO software only if $T_3 > T_1 + T_2$ and by only accepting new versions with increasing expiration time. Importantly, the depositor can always exit the protocol whenever the expiration time of the deployed AO software approaches $T_3$. We develop this topic in more detail in Appendix~\ref{app:ver-lifecycle}.

\subsection{Security Properties}
The TEE architecture provides the following security properties:

\begin{property}[TEE Code Integrity]
    \label{prop:code-integrity}
    Once deployed, the TEE code cannot be modified by an AO operator, AWS administrators, or any other party, except as outlined in sections ~\ref{subsec:TEE-upgrades} and ~\ref{sec:version-expiration}. \\
    
    \textit{Note: We analyze this property in more detail as Guarantee \ref{gar:tee-code-integ} in Sec.\ref{sec:sec-guarantees}.}
\end{property}

\begin{property}[No Pre-Signing] 
    The AO cannot pre-sign any PSBT until after the steps outlined in section ~\ref{sec:tee} and subsections~\ref{UnbondingDisputeResolutionLogic} and~\ref{RebalanceDisputeLogic}. 
\end{property}

\subsection{Attestation and Verification}
\label{sec:attestation}
Attestations are documents produced by the Enclave application to certify that a computation, resolution, or state assessment was performed according to the core logic implemented in a specific EIF image. Each attestation always reports the AO public key, all PCR registries of an Enclave application deployment, and is signed by the AWS Nitro Enclave service to demonstrate such values are legitimate. The attestation document can be verified using the AWS Nitro Enclave Public Key Infrastructure~\cite{nitro-attestation}.

This provides a guarantee to all protocol parties that a specific AO deployment runs a codebase that both the TO and the AO operator agreed upon. The TO would not participate in the Protocol User Setup Ceremony without having verified an attestation that reports the AO public key and the expected PCR registry values. Appendix~\ref{app:repro-build} describes how PCR values and attestation contents define this process.

\section{Security Analysis}
\label{sec:security}
\subsection{Overview}
The BSA protocol achieves a trust model where depositors only need to trust that at least one AO ($1$-of-$k$) or the TO remains honest, achieving $1$-of-$(k+1)$ security (see section ~\ref{subsec:trust-model}). This represents a significant improvement over traditional $m$-of-$n$ threshold schemes that require an honest majority. The security boundary is reached when all AOs and the TO are both corrupted. Additionally, in Lombard's model, the TO is a set of validators, called the Security Consortium, which self-coordinate via a Tendermint~\cite{tendermint} based consensus protocol. 

\subsection{Trust Model: $1$-of-$k$ vs $1$-of-$(k+1)$}
\label{subsec:trust-model}
We use the term $1$-of-$(k+1)$ security to refer to $k$ independent AOs and $1$ TO. The TO is itself a Byzantine Fault Tolerant distributed system running the Tendermint~\cite{tendermint} consensus protocol, progressing correctly if less than one-third of validators are dishonest. The distinction between $1$-of-$k$ and $1$-of-$(k+1)$ refers specifically to the requirement that either $1$-of-$k$ AOs must remain \emph{correct}, or, if all AOs are offline, then the protocol still achieves Protocol Safety~\ref{def:protocol-safety} under the Failure Model~\ref{sec:failure-model} if the TO is honest. If the depositor's security threshold is strictly reliant on AOs, excluding the TO, then we have $1$-of-$k$ security; otherwise, we have $1$-of-$(k+1)$ security. 

\subsection{Security Guarantees}
\label{sec:sec-guarantees}
We now outline the security guarantees of the BSA protocol with respect to the definitions in section ~\ref{sec:security-goals}. The protocol achieves $1$-of-$(k+1)$ security, requiring that at least one of $k$ AOs (see definition~\ref{def:ao-correctness}) is correct or the TO remains honest.
Security guarantees ensure that funds cannot be stolen or lost under the Failure Model~\ref{sec:failure-model}.

\begin{guarantee}[Depositor Safety]
\label{gar:depositor-safety}
    Under the assumption that at least one AO remains correct, a depositor's funds are safe from theft by a malicious TO. Specifically:
    \begin{itemize}
        \item A correct depositor can withdraw their native Bitcoin unilaterally after time lock $T_1$ (if no challenge) or at most $T_1+T_2$ (after challenge resolution).
        \item A malicious TO cannot steal funds through false rebalances ($\text{VA}\rightarrow\text{RCA}$) or unjustified challenges ($\text{UTA}\rightarrow\text{UCA}$), as any correct AO (online and operational) verifies the SAR state and resolves the dispute in favor of the depositor within $T_2$.
        \item Even if an AO keypair is compromised, they cannot steal funds, as AO spending is constrained by emulated covenants to the depositor address only (see section ~\ref{sec:psbt-convenant-emulation}). The most severe action available to a malicious AO operator is to take the AO offline, which has no impact if other correct AOs remain online.
    \end{itemize}    
    See Appendix~\ref{app:depositor-time} for additional remarks.
\end{guarantee}

\begin{guarantee}[Token Operator Safety]
\label{gar:to-safety}
    The TO is safe from malicious depositors and malicious AO operators:
    \begin{itemize}
        \item If a depositor initiates an illegitimate unbonding request ($\text{VA}\rightarrow\text{UTA}$ without updating SAR to \texttt{Withdrawn} or \texttt{Rejected}), the TO can challenge within $T_1$, and funds remain in $\text{UCA}$ until $T_2$ expires, enabling the TO to claim them. All correct AOs verify the SAR state and observe the challenge to be legitimate, allowing the TO to claim funds after $T_2$ expires.
        \item If a depositor's position becomes under-collateralized, the TO can initiate a rebalancing ($\text{VA}\rightarrow\text{RCA}$). Correct AOs verify the destination chain imbalance exists and do not dispute, allowing the TO to claim funds after $T_2$ expires. Incorrect AOs cannot deviate from the prescribed behavior due to property~\ref{prop:code-integrity}.
    \end{itemize}
\end{guarantee}

\begin{guarantee}[Protocol Safety]
\label{gar:protocol-safety}
    The protocol progresses towards a state ensuring Protocol Safety under Guarantees \ref{gar:depositor-safety} and \ref{gar:to-safety}.\\

    That is, Protocol Safety holds if at least one of the following holds: (1) at least one AO is correct, or (2) the TO is correct. If at least one AO is correct, disputes are resolved within time lock $T_2$ through the AO's deterministic dispute resolution logic (sections ~\ref{RebalanceDisputeLogic} and ~\ref{UnbondingDisputeResolutionLogic}). If all AOs are offline but the TO is correct, protocol state transitions occur via Bitcoin script timelocks:  the TO can claim funds from challenge addresses ($\text{UCA}, \text{RCA}$) after $T_2$ expires, ensuring progress to a TO-safe state. The deterministic nature of Bitcoin timelocks ensures progress, regardless of AO availability.
\end{guarantee}

\textit{Note: we do not consider protocol liveness~\ref{def:protocol-liveness} as a property in this section given that we are unconcerned with finality that is reached in a non-protocol-safe manner. However, the reader may observe that eventually, the protocol will always progress to an end state (unilateral control of funds by the depositor or TO) if either the depositor or TO is online.}

\begin{guarantee}[Covenant Enforcement]
\label{gar:covenant-enforcement}
    Even if AWS Nitro Enclaves or KMS are compromised, Bitcoin can only be transferred to either: (1) the depositor's address, or (2) the TO's address (if AO does not act within $T_2$). This is enforced by emulated covenants (section~\ref{sec:psbt-convenant-emulation}): all PSBTs exchanged during the Protocol User Setup Ceremony commit to specific output addresses through pre-signing with fixed outputs. Even if the TEE or KMS is compromised, the pre-signed PSBTs cannot be modified to create new spending paths, as they are cryptographically committed. The only valid spending paths are defined during the original setup ceremony, preventing the establishment of new spending paths to third parties.
\end{guarantee}

\begin{guarantee}[TEE Code Integrity]
\label{gar:tee-code-integ}
    The TEE cannot be modified by AO operators, AWS administrators, or any other party, except through the upgrade mechanism in sections~\ref{subsec:TEE-upgrades} and~\ref{sec:version-expiration}. The TEE enforces code integrity through hardware-enforced isolation. The Enclave Image File (EIF) is cryptographically measured via Platform Configuration Registers (PCRs), and any modifications to the codebase results in different PCR values, which can be verified through attestations (section ~\ref{sec:attestation}). KMS policy restrictions prevent access to signing keys outside the TEE, and the dispute resolution logic (sections ~\ref{RebalanceDisputeLogic} and ~\ref{UnbondingDisputeResolutionLogic}) requires state verification before signing, preventing pre-signing attacks.     
\end{guarantee}

\begin{guarantee}[Protocol Immutability]
\label{gar:protocol-immutability}
    No changes can be made to protocol design (adapter removal ~\ref{sec:perimeter}, SAR upgrades ~\ref{sec:sar-upgrades}, TEE version deprecation ~\ref{subsec:TEE-upgrades}) before the time lock $T_3$ expires. All protocol modifications that impact security goals require time lock $T_3$. During this period, depositors can withdraw their funds unilaterally (see Guarantee ~\ref{gar:depositor-safety}) using previously agreed upon TEE versions, ensuring no party is forced to accept changes they disagree with.
\end{guarantee}

\subsection{Hypothetical Risks with Failure Analysis}
\label{sec:attack-model}
In the failure model (Sec.~\ref{sec:failure-model}) we assume the TO and the depositor to be Byzantine, i.e., they may deviate arbitrarily, by acting dishonestly or maliciously, to maximize their own gain. Throughout the paper, security is analyzed under the assumption that AOs can only fail by being offline, i.e., failing to fulfill time-sensitive duties. We now characterize how the protocol behaves when this assumption is relaxed.
\\ 

The system fails when Protocol Safety~\ref{def:protocol-safety} is violated: either the depositor cannot regain control of their funds (Depositor Safety fails Def.~\ref{def:depositor-safety}), or the TO cannot reclaim collateral when tokens exit the perimeter (Token Operator Safety Def.~\ref{def:token-operator-safety} fails). Failures arise from: (1) AO unavailability or key compromise, or (2) TO quorum corruption or inability to reach consensus.

\subsubsection{AO Failure Types}
\label{subsub:ao-fail-type}
\textbf{Availability (offline).} An AO is \textit{unavailable} if it does not respond within $T_2$ to sign dispute resolutions. If there are offline AOs but at least one correct AO, all guarantees hold. If \textit{all} AOs are offline, the guarantee of Protocol Safety~\ref{gar:protocol-safety} can only be satisfied by an honest TO: a correct TO can claim funds from challenge addresses after $T_2$; a malicious TO can steal depositor funds via false rebalances (violating Depositor Safety~\ref{gar:depositor-safety}) or fail to challenge a depositor's illegitimate unbonding request (violating Token Operator Safety~\ref{def:token-operator-safety}).

\textbf{Key compromise.} An AO's signing key is \emph{compromised} if the key is leaked such that an adversarial actor can use it to execute transactions on behalf of the AO. The leak of one AO key is enough for a malicious depositor to independently perform the following attack: initiate an illegitimate unbond, receive a legitimate challenge from the TO, and sign the resolution in their favor using the leaked key. Depositor Safety~\ref{gar:depositor-safety} holds, and Token Operator Safety~\ref{gar:to-safety} fails. The Guarantee of Covenant Enforcement~\ref{gar:covenant-enforcement} fails in a non-significant way, the depositor can still send funds to anywhere they control. 

\textbf{TEE Failure.} An AO operator can access and update the code deployed in the enclave to something not approved by the TO. In this event the guarantees of TO Safety~\ref{gar:to-safety}, Protocol Safety~\ref{gar:protocol-safety}, TEE Code Integrity~\ref{gar:tee-code-integ}, and Protocol Immutability~\ref{gar:protocol-immutability} (specifically, TEE version deprecation) all fail. Covenant Enforcement~\ref{gar:covenant-enforcement} holds, and Depositor Safety~\ref{gar:depositor-safety} holds if there is at least one correct AO remaining, or if the AO with TEE failure is colluding with said depositor. 

\subsubsection{TO Failure Types}
\textbf{Inability to reach consensus (no sign).} If at least $\nicefrac{1}{3}$ of LSC validators are offline the TO cannot sign. The TO cannot initiate rebalances or challenge depositor unbonding requests. Protocol Safety~\ref{gar:protocol-safety} is degraded. Depositor Safety ~\ref{gar:depositor-safety} is ensured if (1) they have a UTXOs in their VA address, or (2) they do have UTXOs in a challenged state, then they require at least one AO to be correct. Respectively, a depositor (1) can unbond and will not be (illegitimately) challenged, and (2) if a TO illegitimately challenges and then goes offline, the depositor still requires one correct AO to resolve the dispute. Token Operator Safety~\ref{gar:to-safety} is degraded: the TO cannot reclaim funds from challenge addresses if it cannot sign.

\textbf{Quorum corruption.} The TO requires a $\nicefrac{2}{3}$ supermajority to sign. If two-thirds or more of LSC validators are Byzantine, they can form a quorum and produce malicious outcomes, e.g., signing false rebalances, signing false unbond challenges, or refusing to challenge illegitimate unbonding. In this case, Protocol Safety~\ref{def:protocol-safety} is not achieved: a malicious depositor can steal funds if the TO does not challenge illegitimate unbond requests, or a corrupted TO together with all AOs offline can steal depositor funds. Token Operator Safety~\ref{gar:to-safety} fails and Depositor Safety~\ref{gar:depositor-safety} can fail if all AOs are offline (see Sec.~\ref{subsub:ao-fail-type}).

\subsubsection{Guarantee Summary by Failure Type}
Table~\ref{tab:failure-guarantees} summarizes which guarantees remain enabled under each failure scenario. ``Yes'' indicates the guarantee holds; ``No'' indicates it fails. For the table below, unless otherwise specified, we assume the depositor to be Byzantine. 

\begin{table}[H]
\centering
\caption{Guarantees under failure types. Dep. = Depositor Safety~\ref{gar:depositor-safety}; TO = Token Operator Safety~\ref{gar:to-safety}; Safety = Protocol Safety~\ref{gar:protocol-safety}}
\label{tab:failure-guarantees}
\vspace{0.5em}
\scriptsize
\begin{tabular}{llccc}
\toprule
\textbf{Failure} & \textbf{Condition} & Dep. & TO & Safety \\
\midrule
1 AO offline & $\geq 1$ AO correct & Yes & Yes & Yes \\
All AOs offline & TO Honest & Yes & Yes & Yes\\
All AOs offline & TO Malicious & No & Yes & No \\
AO key leak & Dep. Malicious & Yes & No & No\\
TO no consensus & $\geq 1$ AO correct & Yes & No & No \\
TO no consensus & All AOs offline & Yes & No & No \\
TO quorum corruption (non-rational) &  $\geq 1$ AO correct & Yes & No & No\\
TO quorum corruption (non-rational) &  all AOs offline & No & No & No\\
\bottomrule
\end{tabular}
\end{table}

\subsection{Monitoring Requirements}
\label{sec:monitoring-requirements}
Security guarantees depend on monitoring of both the Bitcoin network and the destination chain.

The TO must monitor Bitcoin transactions to detect illegitimate unbonding (VA$\rightarrow$UTA) within $T_1$ and challenge if necessary. Monitoring the Smart Account Registry is necessary to observe legitimate rebalancing events before executing the transition (VA$\rightarrow$RCA) on the Bitcoin blockchain.

Correct AOs must monitor Bitcoin to detect unfair rebalance (VA$\rightarrow$RCA), or unfair challenges (UTA$\rightarrow$UCA) within $T_2$ and resolve the dispute if necessary. Also, by monitoring the SAR, AOs acquire knowledge about new instances of the BSA protocol to safeguard.

Failure to monitor within relevant timelock windows can lead to violation of the security guarantees outlined in section~\ref{sec:sec-guarantees}.

\section{Limitations and Future Work}
\label{sec:limitations}

\subsection{Current Limitations}
\begin{enumerate}
    \item \textbf{Infrastructure Integrity:} The protocol trusts AWS Nitro and KMS infrastructure for the AO role. While this is a reasonable assumption, it introduces a dependency on a single technology and service provider. We view this risk as acceptable, as in the event of a full compromise of AWS KMS keys, the architectural design choices limit the only possible action to be the unbonding of Bitcoin in favor of the depositor.
    
    \item \textbf{Timelock Trade-offs:} The timelock periods $T_1$ and $T_2$ create a trade-off between security and user experience. Longer time locks provide greater security but also introduce potential delays for depositor unbonds and TO rebalances. 
    
    \item \textbf{Liquidation Granularity:} Model 1 (Collaborative Rebalance, Sec.~\ref{sec:coll-rebalancing}) requires user availability to create new PSBTs, which may not be ideal for all use cases. Model 2 (UTXO-Based Rebalancing, Sec.~\ref{sec:utxo-rebalancing}) may impede user experience by requiring multiple signing of PSBTs during the setup ceremony (Sec. \ref{subsubsec:setup-cermony}).

    \item \textbf{Downstream Oracle Failure:} The BSA protocol does not mitigate risks arising from downstream DeFi protocol price oracle failures or manipulations, which would lead to unfair rebalances, as the perimeter only tracks on-chain balances.
\end{enumerate}

\subsection{Future Work}

\begin{enumerate}
    \item \textbf{Resynchronization Dependency}: Supporting multiple sources beyond the Lombard Security Consortium (e.g., other trusted sources and services) to provide trusted check points for the destination chains—used during resynchronization after extended AO downtime (see section~\ref{subsec:light-client-sync})—would improve AO liveness.
    
    \item \textbf{Multi-Client Destination Chain Architecture:} Implement a selection of destination chain clients and consensus between them to leverage code diversity and expand the set of supported destination chains.
    
    \item \textbf{Bitcoin Protocol Upgrades:} If Bitcoin adopts native covenant opcodes (e.g., \texttt{OP\_CAT}~\cite{bip347}, \texttt{OP\_CHECKTEMPLATEVERIFY}~\cite{bip119}), the protocol could be simplified and efficiency improved.
    
    \item \textbf{Alternative TEE Providers:} Explore support for TEE providers beyond AWS Nitro to reduce infrastructure dependency on both the cloud provider (e.g., Google Cloud Platform, Microsoft Azure) and CPU manufacturer (AMD SEV-SNP, Intel TDX).
\end{enumerate}

\section{Conclusion}
\label{sec:conclusion}

This paper introduced Bitcoin Smart Accounts (BSA), a novel protocol that enables native Bitcoin to access DeFi applications while maintaining self-custody and minimizing trust assumptions. We achieve this through the combination of emulated Bitcoin covenants (using PSBTs and Taproot scripts), a TEE-based arbitration system, and smart contracts that enable DeFi platforms to operate without necessitating protocol-level modifications.

BSA achieves a trust model where depositors only need to trust that at least one Arbitration Oracle or the Token Operator remains honest, achieving $1$-of-$(k+1)$ security, a significant improvement over traditional $m$-of-$n$ threshold schemes. A single honest AO can counter any decision made by the TO, providing strong safety guarantees for depositors.

We provided security evaluations demonstrating correctness, safety, and liveness properties under our trust model. Our design enables native Bitcoin to serve as collateral in lending markets and other DeFi protocols without requiring users to relinquish custody of their funds, mimicking a tri-party agreement.

The protocol represents a significant step toward unbonding Bitcoin's vast capital for DeFi applications while preserving Bitcoin's core values of self-custody and trust minimization. We believe that Bitcoin Smart Accounts provide a practical and secure foundation for the growing Bitcoin DeFi ecosystem. 

\appendix
\section*{Appendices}
\section{Bitcoin Taproot Address Tweaking}
\label{app:address-tweaking}

A Taproot address commits to an \emph{internal key} $P$ and a list of spending scripts, organized in a Merkle tree, via its Merkle root $m$~\cite{taproot}, by combining them in a single \emph{tweaked key}. Such a \emph{tweaked key} $Q$ is included in the \texttt{ScriptPubKey} that yields the Taproot address, and it is computed as follows: $Q = P + tG$ where $t = \mathsf{hash}(p \| m)$, $p$ is the 32-byte encoding of the $x$-coordinate of point $P$ and the point $G$ is the generator point of the \texttt{secp256k1} elliptic curve. Any variation either of the internal key, or of the scripts list produces a different \texttt{ScriptPubKey}, so a different taproot address. For full specifications, please refer to BIP-341~\cite{taproot}.

\subsection{Nothing Up My Sleeves (NUMS) Internal Key}
\label{app:nums}
All BSA Taproot addresses use a \textit{Nothing Up My Sleeves} (NUMS) point as the internal key $P$~\cite{taproot}. A NUMS point is an elliptic curve point assumed to have an unknown discrete logarithm with respect to the \texttt{secp256k1} generator point $G$, implying it is computationally infeasible to use it to satisfy a signature verification algorithm. Additionally, NUMS keys can be constructed deterministically starting from a known set of inputs so that it is verifiable that no party can have chosen it to embed a backdoor. BIP-341~\cite{taproot} describes as an example the NUMS point, 

\begin{equation}
    H = \mathsf{lift\_x}(\mathtt{0x50929b74c1a04954b78b4b6035e97a5e078a5a0f28ec96d547bfee9ace803ac0})
\end{equation}

constructed by taking the hash of the standard uncompressed encoding of $G$ and using the result as the $x$-coordinate. A number of NUMS keys can be generated starting from $H$ by taking $H + rG$ where $r$ is any integer smaller than the order of the \texttt{secp256k1} curve.

Since a valid taproot address includes a key spending path, BSA disables it using a NUMS construction, resulting in only script path spending being enabled, used for multi-signature in covenant emulation and time-lock conditions. 

BSA leverages the following NUMS construction to commit protocol-specific information in each protocol addresses. Each NUMS point is computed following the BIP-341~\cite{taproot} suggestion $H +rG$ by taking $r = sha256(\text{address type} || \text{tweak data}) \bmod n$, where $n$ is the order of the \texttt{secp256k1} curve. The \emph{address type} is a simple label identifying which of the four address types should be generated (VA, UTA, UCA, RCA), while the \emph{tweak data} includes all the information that characterizes the protocol instance such as the keys of all involved parties, all timelocks, and the destination address information.

\section{Transaction Reference Table}
\label{app:tx-reference}

Table~\ref{tab:tx-reference} lists the essential Bitcoin state transitions. Abbreviations \textit{Dep}, \textit{TO}, \textit{AO}: Depositor, Token Operator, Arbitration Oracle. \textit{Creator}: creates and signs the PSBT; \textit{Stored}: where it is held (Dep, TO, SAR); \textit{Executor}: signs, finalizes, and broadcasts.

\begin{table}[H]
\centering
\caption{Transaction reference.}
\vspace{0.1618cm}
\label{tab:tx-reference}
\scriptsize
\setlength{\tabcolsep}{6pt}
\renewcommand{\arraystretch}{2}
\makebox[\textwidth][c]{%
\begin{tabular}{>{\raggedright\arraybackslash}p{2.5cm}lcp{1.9cm}@{\hspace{-10pt}}cc>{\raggedright\arraybackslash}p{2cm}>{\raggedright\arraybackslash}p{3cm}}
\toprule
 \textbf{Transaction}  &\textbf{Path}& \textbf{Signers}& \textbf{Creator}& \textbf{Stored} & \textbf{Executor}& \textbf{Fee} & \textbf{Motivation} \\
\midrule
 Unbond request  &VA $\rightarrow$ UTA& TO \& Dep & TO& SAR & Dep & Anchor; CPFP~\cite{cpfp} & Dep initiates withdraw request \\
 Unbond finalize &UTA $\rightarrow$ Dep& Dep & ---& --- & Dep & Dep's UTXOs & Unbond request not challenged, claimed after $T_1$ \\
 Unbond challenge  &UTA $\rightarrow$ UCA& Dep \& TO & Dep& TO & TO & Anchor; CPFP & TO challenges unbond request within $T_1$ \\
 Unbond resolve&UCA $\rightarrow$ Dep& Dep \& AO & Dep& SAR & AO & SIGHASH~\cite{taproot} & AO sides with depositor within $T_2$ \\
 Unbond resolve (expired)&UCA $\rightarrow$ TO& TO & ---& --- & TO & TO UTXOs & TO claims after $T_2$; challenge was valid \\
 Rebalance request  &VA $\rightarrow$ RCA& Dep \& TO & Dep& TO & TO & Anchor; CPFP & TO initiates rebalance request \\
 Rebalance resolve&RCA $\rightarrow$ Dep& Dep \& AO& Dep& SAR & AO & SIGHASH & AO sides with Dep; unfair rebalance \\
 Rebalance resolve (expired)&RCA $\rightarrow$ TO&  TO & ---& --- & TO & TO UTXOs & TO claims after $T_2$; rebalance was valid \\
\bottomrule
\end{tabular}%
}
\end{table}

\textit{Note: in the above table, the terms \textit{Creator} and \textit{Executor} are as defined in section ~\ref{sec:psbt-convenant-emulation}. The Executor column indicates the recommended party, though, after both parties have signed, either signer may execute the state transition. Storage by the TO refers to Lombard Ledger, the Security Consortium's ledger.}

\section{Reproducible Build and Outcome for Arbitration Oracle}
\label{app:repro-build}
An important desired characteristic of Trusted Execution Environment (TEE) is the ability to attest, or provide proof, that the outcome of a computation from within the TEE is the result of the execution of a specific codebase with a corresponding set of inputs and states. AWS Nitro Enclaves provides this feature via \emph{attestations}: documents signed-off by the Nitro hardware and software, validated via the AWS Public Key Infrastructure, which report measurements of the runtime state inside a running enclave. In particular, there are two values, or Platform Configuration Registries (PCR), that are used to verify the AO operation correctness:
\begin{enumerate}
    \item \textbf{PCR0}: reports the sha384 digest of the Enclave Image File running in the enclave, effectively certifying that a specific codebase was built and assembled in a determined runtime.
    \item \textbf{PCR8}: reports the sha384 digest of the certificate that signed the Enclave Image File running in the enclave, certifying that a deployed artifact has been signed by the TO.
\end{enumerate}

While PCR8 is a simple certificate-match verification, the reliability of PCR0 depends on the ability to reproduce the same Enclave Image File from the same source code. To achieve this, the AO implementation relies on a deterministic and reproducible toolchain, which includes:
\begin{itemize}
    \item A deterministic Rust compiler,
    \item Hash-locked dependencies for libraries needed at compile time to always produce the same statically linked binary,
    \item Hash-locked dependencies for tools needed at execution time, i.e. network proxies,
    \item Pinned base image to construct the enclave runtime,
    \item Pinned Nitro Enclave toolchain to pack together the Enclave Image File from the runtime definition and the AO logic build artifacts.
\end{itemize}

\subsection{AO Attestation Contents}

Beyond the PCR measurements included in all AWS Nitro Enclave attestations, such documents offer the possibility to include additional application-specific data. The AO implementation always sets for every attestation:
\begin{itemize}
    \item The Bitcoin public key of the AO to correctly identify the attester identity,
    \item The latest Ethereum finalized checkpoint to assess what Ethereum state the AO decision was based on,
    \item The hash of the AO resolution and the Bitcoin input data used to derive this resolution.
\end{itemize}

\section{Arbitration Oracle Key Lifecycle}
\label{app:arb-key}
The correctness of the AO running in the TEE is directly connected to the security of the key pair representing its Bitcoin identity. The main challenge to overcome in having a TEE operating a Bitcoin identity is the lack of persistent storage to rely on in case of restarts of the AO-deployed artifact. In fact, once a keypair is securely generated in an enclave using a crypto secure pseudo-random number generator, they are used in the Bitcoin scripts, addresses, and PSBTs reported in Sec.~\ref{sec:bitcoin-arch} which remains valid for an indefinite amount of time. However, a restart of the AO instance would lose any previous state, thus permanently losing the previous keypair and ostensibly implying it could only generate a brand new identity. This issue is overcome using AWS KMS and its native integration with AWS Nitro Enclaves attestations.

We report below all the operations an AO enclave and the operator deploying its logic in a cloud instance (referred to as the \emph{AO operator}) carry out to achieve a secure persistent usage of a Bitcoin identity in case of restarts, upgrades, or crash failures of the AO TEE logic.

As a brief summary of the solution (explored in the proceeding section), we can say that the enclave logic uses AWS KMS to encrypt its Bitcoin private key. Then, it creates an attestation using the Nitro Secure Module (NSM) that new instantiations of the enclave can use to verify that the decoded identity originates from a legitimate execution of an AO in an AWS Nitro Enclave. Such NSM attestations are also validated by AWS KMS, which only decrypts the encrypted backup of the Bitcoin private key to the authorized software.

\subsection{Key Initialization}
\label{app:kms-key-initialization}
On initial boot, the AO running in the enclave has no identity set, and it receives from the AO operator the \texttt{Init} command which makes it:
\begin{enumerate}
    \item Generate a Bitcoin keypair (secp256k1) requesting a source of randomness from the Nitro Secure Module. Additional and independent sources of randomness can be fed to contribute to this process.
    \item Request from AWS KMS the creation of a new key with encrypt and decrypt capabilities configured with the constraints described in Sec.~\ref{app:kms-key-constraints}.
    \item Encrypt the Bitcoin secret key using the created AWS KMS key (locally in the enclave or via secure communication with AWS KMS)
    \item Produce, using the Nitro Secure Module available inside the enclave, an attestation that includes, in the user data field, the hash of the concatenation of the following data: the newly generated Bitcoin public key, the encrypted private key, and the KMS Key ID.
    \item Return the attestation to the cloud operator along with the information hashed in the user data.
\end{enumerate}

At the end of a successful initialization process, the AO operator will obtain:
\begin{itemize}
    \item A Nitro Enclave initialized with the logic of an AO and a legitimate identity suitable to play the AO role in BSA protocol instances.
    \item The Bitcoin public key of the AO to share to depositors for inclusion in the Protocol User Setup Ceremony.
    \item An encrypted version of the Bitcoin private key corresponding to the public key of the identity of the AO.
    \item The KMS Key ID used to encrypt the AO Bitcoin private key.
    \item An AWS Nitro Enclave attestation that the AO Bitcoin keypair has been generated according to the logic of a specific version of the AO (PCR0 measurement) released and signed off by the token operator (PCR8 which includes the hash of the TO certificate)
\end{itemize}

\subsubsection{Key Restore}
After a restart of the AO enclave, the cloud operator can restore a previously generated AO identity by sending to the TEE a Restore command and the information obtained during the first initialization: the Bitcoin public key, the encrypted secret key, the KMS Key ID, and the attestation committing to such information. The AO implementation running inside the Nitro Enclave:
\begin{enumerate}
    \item Verifies the attestation using the AWS Nitro Enclave root certificate publicly provided by AWS \cite{nitro-attestation} and embedded in the enclave image.
    \item Decrypts the Bitcoin secret key using the AWS KMS Key referenced by the input Key ID.
    \item Verifies the decrypted secret key corresponds to the Bitcoin public key passed as input.
\end{enumerate}

\subsubsection{AWS KMS Key constraints}
\label{app:kms-key-constraints}
Constraints on the key usage are expressed in AWS KMS via a usage policy. The core of the AWS Nitro Enclave integration with AWS KMS is the possibility to specify in the key usage policy constraints based on the PCR measurements of the Nitro Secure Module available in the Nitro Enclave. In the AO case, the KMS Key created during the initialization phase (Sec.~\ref{app:kms-key-initialization}), includes a condition on the PCR8 measurement. Such registry is set to the hash of the certificate released by the entity that signed the Enclave Image File running inside the TEE. By setting such a value to a certificate in control of the Lombard Security Consortium, the AO operator can only restore a legitimate identity inside versions of the enclave software released and signed-off by the Consortium. KMS will not authorize any other usage of the decryption key.

Additionally, the key policy explicitly denies updates of the policy itself, locking out the AO operator or any root admin of the AWS Organization.

\section{AO Availability Requirement Estimation}
\label{app:avail-req}
The guarantee of Depositor Safety (Sec.~\ref{gar:depositor-safety}) pivots on the assumption the AO can protect a depositor from potential unfair challenges by the TO. This section aims to outline the availability requirements for an AO operator to fulfill the $1$-of-$k$ requirement in the BSA protocol.

\subsection{Protocol Parameter Constraints}
An AO will run software authorized by the TO if it has an expiration time of at least $T_3$ (Sec.~\ref{sec:protocol-parameters}), which is the minimum time window any protocol change needs to take effect. The relation $T_3 > T_1 + T_2$, introduced in Sec.~\ref{sec:protocol-parameters}, implies any depositor, in order to be sure to leverage the protection of an online AO, has a time window $\delta$ to judge if their participation in the BSA protocol should come to an end (e.g., by withdrawing funds with an unbonding operation) if the new protocol conditions are not satisfactory, where, 
\begin{equation}
    \label{eq:delta-exit}
    \delta = T_3 - (T_1 + T_2)
\end{equation}
Importantly, time parameter $T_3$ is mutable, whereas, for an instance of the protocol, time parameters $T_1$ and $T_2$ are immutable: they are set in the locking scripts committed by addresses~\ref{sec:script-addresses}.

In order to keep the aforementioned relation valid, there should always be a timeframe within any $T_3$ window set by the TO where an AO can correctly validate, and, if needed, resolve challenges by the TO: depositor-challenged unbonding requests or rebalancing requests. The minimum amount of time it takes for a malicious TO to challenge a correct unbond and unfairly collect depositor funds is $T_2$. In fact, the unbonding challenge transaction by the TO could be mined in the same block as the unbonding request, de-facto skipping the $T_1$ wait time, i.e., the $T_1$ window is intended singularly for TO security, and does not impact AO availability requirements. Therefore, on rebalancing or challenged-unbonding requests $T_2$ is the only time parameter to take into account. Also, we include in our considerations the time $t_{\text{op}}$ needed by the AO to process a challenge, i.e., the time it takes for the AO to formulate a decision using the predefined logic outlined in \ref{sec:tee}, and ensure, if required, that any action is correctly applied on the Bitcoin blockchain. Thus, an initial constraint about the AO availability $\text{uptime}'_{AO}$ is:

\begin{equation}
\label{eq:ao-uptime}
    \text{uptime}'_{AO} > \frac{T_3-(T_2-t_{\text{op}})}{T_3}
\end{equation}

where it is an implicit requirement that $t_{\text{op}} < T_2$ otherwise the AO duties could never be fulfilled, and that $\text{uptime}'$ is calculated per $T_3$. This ensures that for any point in time $t$ within a $T_3$ window from deployment by a TO, the AO must be online for at least $t_\text{op}$ within the interval $[t, t + T_2]$.

\subsection{Ethereum Consensus Constraints}
\label{app:eth-consensus}
The global timelock for the TO to change protocol rules $T_3$ is not the only critical protocol parameter BSA parties should monitor to ensure safety guarantees hold. In fact, the AO software running in the enclave relies on an Ethereum light client to validate protocol states from the Smart Account Registry and resolve challenges accordingly. This means that the inability to produce a reliable view of the Ethereum blockchain is the same as preventing an AO from fulfilling their arbitration tasks. 

An AO can access the Ethereum state in a reliable way by initializing its light client (running inside the TEE) with a trusted checkpoint obtainable in two possible ways: from the TO by taking a checkpoint signed-off by Lombard's Security Consortium (LSC), or from a previous attested checkpoint produced by the AO.

Independently from the source, according to the Ethereum consensus specification~\cite{wsp}, each finalized checkpoint can be considered a reliable view of the network state from the moment it is finalized for a time period called the \emph{Weak Subjectivity Period} (WSP). After the WSP expires, the given checkpoint may expose the client to a class of attacks from a minority of validators. To prevent that, an offline AO could only restart using a trusted checkpoint not older than the current time minus a default minimum WSP. This mechanic impacts the uptime requirements of the AO.

At initial startup, an AO is deployed using an LSC trusted checkpoint, which can be validated in an independent manner by the AO operator before use. After initial startup, the AO can autonomously produce a new trusted checkpoint while it is operational. Thus, a requirement on the AO uptime is derived by the need to be online at least once during the WSP of the latest available trusted checkpoint, for an amount of time $t_{\text{check}}$, defined as the time it takes to synchronize with the network and produce an attestation of a finalized checkpoint.

\begin{equation}
\label{eq:ao-uptime-eth}
    \text{uptime}''_{AO} > \frac{t_{\text{check}}}{\text{WSP}}
\end{equation}

In conclusion, the complete availability requirement for an AO to fulfill their $1$-of-$k$ duty, $\text{uptime}_{AO}$ is:

\begin{equation}
\label{eq:ao-uptime-final}
    \text{uptime}_{AO} = \text{max}\{\text{uptime}'_{AO}, \text{uptime}''_{AO}\}
\end{equation}

\section{AO Version Lifecycle}
\label{app:ver-lifecycle}
The AO operator is responsible for verifying accepted software versions posted by the TO have an expiry such that $\delta = T_3 - (T_1 + T_2) > 0$, ensuring depositor safety, see Guarantee~\ref{gar:depositor-safety}. However, the TO is not constrained to ensure this relation holds for software version expiration. In fact, this relation is only enforced by the SAR where $T_3$ reflects SAR upgrades and adapter removal, while the TO, may set an arbitrary expiration for any new AO version.

We outline below safety measures taken by the AO to ensure (1) initialization is done according to protocol specifications and (2) an optimal upgrade choice is taken by AOs when accepting a new version posted by the TO.

\subsection{AO Version Deployment Criterion}
The AO operator finds the current parameter $T_3$ on the SAR. We consider software version expiration times $t_{\text{exp}}$ as valid if $t_{\text{exp}} - t_{\text{present}} > T_3$. 

\subsection{AO Version Upgrade Criterion}
Assume the current accepted version by an AO has expiry time $t_{\text{exp}}$. A new version posted by the TO with expiry time $t_{\text{exp}}'$ will be accepted by the AO if,
\begin{equation}
    t_{\text{exp}}' > t_{\text{exp}}
\end{equation}

\subsection{Depositor Time-Based Decision Criterion}
\label{app:depositor-time}
If the depositor wishes to join a protocol with a specific online AO, they must ensure that 
\begin{enumerate}
    \item $T_1 + T_2 + T_{\text{op}} < T_3$ from SAR
    \item $t_{\text{present}} + T_1 + T_2 + T_{\text{op}} < t_{\text{exp}}$ for AO expiration
\end{enumerate}
where $T_{\text{op}}$ is the maximal operational time it takes for the depositor to execute an unbonding transaction and have this confirmed on the Bitcoin network. \newline

If the AO is online, and a depositor wishes to exit the protocol the depositor can only safely exit, provided the AO remains online, confirming an unbonding on Bitcoin before $t_{\text{exp}} - (T_1+T_2)$.

\end{document}